\documentclass[a4paper,fleqn]{cas-dc}

\usepackage[numbers]{natbib}

\usepackage{glossaries}
\usepackage{listings}
\usepackage{xcolor}

\usepackage{multirow}
\usepackage{colortbl}
\usepackage{url}

\usepackage{circledsteps}
\newcommand{\blackCircle}[1]{\Circled[inner color=white, outer color=black, fill color=black]{#1}}

\def\tsc#1{\csdef{#1}{\textsc{\lowercase{#1}}\xspace}}
\tsc{WGM}
\tsc{QE}
\tsc{EP}
\tsc{PMS}
\tsc{BEC}
\tsc{DE}
\begin{document}

\newacronym{IoT}{IoT}{Internet of Things}
\newacronym{RoT}{RoT}{Root of Trust}
\newacronym{TCB}{TCB}{Trusted Computing Base}
\newacronym{TPM}{TPM}{Trusted Platform Module}
\newacronym{TEE}{TEE}{Trusted Executed Environment}
\newacronym{RA}{RA}{Remote Attestation}
\newacronym{TOCTOU}{TOCTOU}{Time-Of-Check Time-Of-Use}
\newacronym{MPU}{MPU}{Memory Protection Unit}
\newacronym{CPU}{CPU}{Central Processing Unit}
\newacronym{PACBTI}{PACBTI}{Pointer Authentication and Branch Target Identification}
\newacronym{ROP}{ROP}{Return-Oriented Programming}
\newacronym{JOP}{JOP}{Jump-Oriented Programming}
\newacronym{CFI}{CFI}{Control-Flow Integrity}
\newacronym{CFG}{CFG}{Control-Flow Graph}
\newacronym{SROP}{SROP}{Sigreturn Oriented Programming}
\newacronym{ASLR}{ASLR}{Address Space Layout Randomization}
\newacronym{NX}{NX}{Non-Execute}
\newacronym{CFA}{CFA}{Control-Flow Attestation}
\newacronym{QARMA}{QARMA}{Qualcomm ARM Authenticator}
\newacronym{PA}{PA}{Pointer Authentication}
\newacronym{PAC}{PAC}{Pointer Authentication Code}
\newacronym{BTI}{BTI}{Branch Target Identification}
\newacronym{CPI}{CPI}{Code-Pointer Integrity}
\newacronym{FOP}{FOP}{Function Oriented Programming}
\newacronym{ISA}{ISA}{Instruction Set Architecture}
\newacronym{TF-M}{TF-M}{Trusted Firmware-M}
\newacronym{PSA}{PSA}{Platform Security Architecture}
\newacronym{NSPE}{NSPE}{Non-secure Processing Environment}
\newacronym{SPE}{SPE}{Secure Processing Environment}
\newacronym{RTOS}{RTOS}{Real-time Operating System}
\newacronym{API}{API}{Application Programming Interface}
\newacronym{SFN}{SFN}{Secure Function}
\newacronym{IPC}{IPC}{Inter-Process Communication}
\newacronym{SPM}{SPM}{Secure Partition Manager}
\newacronym{FLIH}{FLIH}{First-Level Interrupt Handling}
\newacronym{SLIH}{SLIH}{Second-Level Interrupt Handling}
\newacronym{ITS}{ITS}{Internal Trusted Storage}
\newacronym{HAL}{HAL}{Hardware Abstraction Layer}
\newacronym{IP}{IP}{Intellectual Property}
\newacronym{PoC}{PoC}{Proof of Concept}
\newacronym{DEP}{DEP}{Data Execution Prevention}
\newacronym{OS}{OS}{Operating System}
\newacronym{TFM}{TFM}{TrustedFirmware-M}

\let\WriteBookmarks\relax
\def\floatpagepagefraction{1}
\def\textpagefraction{.001}
\shorttitle{RunPBA - Runtime attestation for microcontrollers with PACBTI}
\shortauthors{Cirne et~al.}

\title [mode = title]{RunPBA - Runtime attestation for microcontrollers with PACBTI}                      
\tnotemark[1]

\tnotetext[1]{This work was supported by Fundação para a Ciência e Tecnologia (FCT), Portugal - 2021.08587.BD.}

%\tnotetext[2]{The second title footnote which is a longer text matter
%   to fill through the whole text width and overflow into
%   another line in the footnotes area of the first page.}

\author[1,2]{André Cirne}[type=editor,
                        %auid=000,bioid=1,
                        %prefix=Sir,
                        %role=Researcher,
                        orcid=0000-0002-3433-9809]
\cormark[1]
\fnmark[1]
\ead{andre.cirne@fc.up.pt}
%\ead[url]{www.jkkrishnan.in}

%\credit{Conceptualization of this study, Methodology, Software}

%\address[1]{, Street 129, 1043 NX Amsterdam, The Netherlands}
\affiliation[1]{organization={Departamento de Ciência de Computadores da Faculdade de Ciências da Universidade do Porto},
                addressline={Rua do Campo Alegre s/n}, 
                city={Porto},
%               citysep={}, % Uncomment if no comma needed between city and postcode
                postcode={4169-007}, 
                %state={Kerala},
                country={Portugal}}

\affiliation[2]{organization={INESC TEC},
                addressline={Campus da Faculdade de Engenharia da Universidade do Porto, Rua Dr. Roberto Frias}, 
                postcode={4200-465}, 
                %postcodesep={}, 
                city={Porto},
                country={Portugal}}

\affiliation[3]{organization={TekPrivacy, Lda},
                addressline={Rua Alfredo Allen n.º 455}, 
                postcode={4200-135}, 
                %postcodesep={}, 
                city={Porto},
                country={Portugal}}

\affiliation[4]{organization={INSIGHTSEC, Lda},
                %addressline={Campus da Faculdade de Engenharia da Universidade do Porto, Rua Dr. Roberto Frias}, 
                %postcode={4200-465}, 
                %postcodesep={}, 
                city={Aveiro},
                country={Portugal}}

\author[4]{Patrícia R. Sousa}[]

\author[1,2]{João S. Resende}[]

\author[1,3]{Luís Antunes}[]

\cortext[cor1]{Corresponding author}

\iffalse
\begin{abstract}
This template helps you to create a properly formatted \LaTeX\ manuscript.

\noindent\texttt{\textbackslash begin{abstract}} \dots 
\texttt{\textbackslash end{abstract}} and
\verb+\begin{keyword}+ \verb+...+ \verb+\end{keyword}+ 
which
contain the abstract and keywords respectively. 

\noindent Each keyword shall be separated by a \verb+\sep+ command.
\end{abstract}

\fi 

%\includegraphics{figs/cas-grabs.pdf}

\begin{highlights}
\item Classification of potential attacks against PACBTI and analysis of how this extension can be used to mitigate control-flow attacks.
\item \textit{RunPBA}, an implementation of a runtime attestation system based on PACBTI to enforce control-flow integrity for embedded devices.
\end{highlights}

\begin{keywords}
embedded systems \sep attestation \sep control-flow attestation \sep runtime attestation \sep PACBTI
\end{keywords}

\maketitle

%-------------------------------------------------------------------------------
\begin{abstract}
The widespread adoption of embedded systems has led to their deployment in critical real-world applications, making them attractive targets for malicious actors. These devices face unique challenges in mitigating vulnerabilities due to intrinsic constraints, such as low energy consumption requirements and limited computational resources.

This paper presents \textit{RunPBA}, a hardware-based runtime attestation system designed to defend against control flow attacks while maintaining minimal performance overhead and adhering to strict power consumption constraints. \textit{RunPBA} leverages \acrfull{PACBTI}, a new processor extension tailored for the Arm Cortex M processor family, allowing robust protection without requiring hardware modifications, a limitation present in similar solutions. 

We implemented a proof-of-concept and evaluated it using two benchmark suites. Experimental results indicate that \textit{RunPBA} imposes a geometric mean performance overhead of only 1\% and 4.7\% across the benchmarks, underscoring its efficiency and suitability for real-world deployment.
\end{abstract}

%-------------------------------------------------------------------------------
\section{Introduction}
%-------------------------------------------------------------------------------

Embedded systems play a pivotal role in our lives. They perform highly critical real-world tasks, like controlling dams, our cars, and  gas pipelines, making them a prime target for attackers. Therefore, the security of these embedded systems must be a priority. Unfortunately, many are developed with constraints that make achieving a secure device challenging.  Most of them aim to be low-cost and energy-efficient, which conflicts with many of the security features we would like to include in these devices.  As is often the case in security, it is necessary to find a balance between the two. 

Simultaneously, the demand for trustworthiness in computing systems has escalated to combat security threats. Attestation is one of the requirements to achieve trust, since it provides evidence to a party about the current configuration and state of a device. Without attestation, it would be impossible for a third party to trust a specific device, as they would not be able to verify the device's state. This feature forms the basis for the implementation of many other processes. Regardless, a challenge remains, which affects many implementations: the \gls{TOCTOU} problem. This issue involves a temporal gap between the measurements of the device state. If not addressed, an attacker could compromise the device and then erase any trace of itself within these intervals, leaving the attesting party unaware of the security breach~\cite{Ankergaard2021,Kuang2022,Makhdoom2023}.

Runtime attestation addresses this issue directly by verifying that software operates under controlled conditions~\cite{Abera2016,Kuang2020}.  However, these solutions come with notable drawbacks. Adding additional checks introduces overhead, making them unsuitable for real-time operations and increasing energy consumption~\cite{Burow2017}. This challenge led to the development of hardware-based systems using custom hardware~\cite{Dessouky2017}. Although these schemes address the earlier issues, they introduce new challenges, such as added complexity and increased costs due to the necessity of custom hardware.

Recent enhancements to the ARMv8-M architecture introduced the \gls{PACBTI}~\cite{Mujumdar2021} extension. This extension mitigates control-flow attacks, similar to those targeted by \gls{CFI} mechanisms. Unlike previous hardware-assisted \gls{CFI} solutions, the \gls{PACBTI} extension is an existing feature already integrated into some processors. Consequently, it has the potential to transform \gls{CFI} into low-end devices, enabling the development of new, more accessible solutions based on this extension, which are easier to adopt than current alternatives.

This paper presents \textit{Runtime PACBTI Attestation (RunPBA)}: a runtime attestation system that takes advantage of the \gls{PACBTI} extension and ARM TrustZone to address the limitations of runtime attestation solutions. RunPBA uses the \gls{PACBTI} extension to monitor for any attempt to escape the intended execution flow. The rest of the \textit{RunPBA} system leverages \gls{TF-M}, a \gls{TEE} for ARM TrustZone, as a \gls{RoT} to securely execute the \textit{RunPBA} logic. We implemented a proof-of-concept of this system and used it to perform real-world tests, where it incurs an overhead of 1.1\% and 4.7\% on two benchmark suites.

In summary, our main contributions are as follows.
\begin{itemize}
    \item \textit{Analysis:} Classification of potential attacks against PACBTI and analysis of how this extension can be used to mitigate control-flow attacks (Section \ref{pacbti-security-risks}).
    \item \textit{Design:}  A runtime attestation system that is able to detect attempts of control-flow attack and properly express them during attestation of a device (Section \ref{runpba}). 
    \item \textit{Implementation:} \textit{RunPBA} - an implementation of a runtime attestation system based on PACBTI to enforce control-flow integrity.(Subsection \ref{runpba_implementation}).
    \item \textit{Evaluation:} Analysis of \textit{RunPBA} showing an overhead of 1.1\% on Coremark PRO benchmark\cite{EEMBC_CoreMark_PRO_2024} and 4.7\% on BEEBS benchmark\cite{Pallister2013}, and a comparison with other \gls{CFI} solutions aimed at the ARM Cortex M family of processors (Section \ref{evaluation}). 
\end{itemize}

The source code of \textit{RunPBA} is available at: \url{https://github.com/MrSuicideParrot/runpba}.

\section{Background knowledge}\label{background-knowledge}
In this study, we draw on various advanced technologies and frameworks. The following section offers a thorough overview of these key concepts, establishing a solid foundation for the detailed analysis and findings discussed in the subsequent sections of this paper. 

\gls{TF-M}~\cite{ArmLimited2024} is an implementation of the \gls{PSA} \gls{IoT} Security Framework for the ARM Cortex-M family of processors that establishes the foundation for a device's \gls{RoT}. Among other features, \gls{TF-M} provides mechanisms to isolate secure and non-secure resources, manage keys, and attest the device. On its core, there is the separation between two environments: the \gls{NSPE} and the \gls{SPE}.  The \gls{NSPE} typically hosts user applications and an \gls{RTOS}, while the \gls{SPE} contains \gls{TF-M} services, with isolation even among services within the \gls{SPE}. This process is normally built exploiting the ARM TrustZone  and \glspl{MPU}.

The \gls{TF-M} consists of three main features: secure booting into both the \gls{NSPE} and \gls{SPE}; enforcing isolation between the two environments while allowing secure communications via \gls{PSA}-defined \gls{API}s; and providing a set of security primitives within the \gls{SPE}~\cite{ArmLimited2024}. 

The \gls{NSPE} composition is delegated to the developer, but is common to have a \gls{RTOS} managing user applications. On the other hand, \gls{SPE} is responsible for securely bootstrapping the device, therefore, it has a stricter composition. The  \textit{\gls{TF-M} Core} implements the inter-process communication, interrupt handling, scheduling, and manage partitions inside the \gls{SPE}. A partition in the \gls{TF-M} is a logical unit that contains a set of services.

The services provided by partitions are known as \gls{RoT} Services, as they act as a \gls{RoT} for \gls{NSPE} applications. Some of these services are mandatory according to the \gls{PSA} \gls{IoT} Security Framework and are referred to as Plataform \gls{RoT}. Others can be developed by the user and are known as Application \gls{RoT}. The isolation level between these components depends on the device configuration.

The \gls{PSA} Security Framework also the different steps a device should have through its lifecycle, along with its security implications. During manufacturing, the device is provisioned which includes the installation of the Platform \gls{RoT} data, its secrets, and enrollment in a device management system. After this phase, the device is locked into a secure state, meaning that secure boot is enabled and any potentially intrusive debugging features, such as hardware debug ports, are disabled. Decommissioned is the final state of the device lifecycle. In this state, the device is considered insecure and its secrets should be permanently inaccessible. The device may only leave this state if a full factory reset is performed, making it indistinguishable from a new device. Regardless, these security states and their characteristics will depend on the device's implementation. Furthermore, additional states or sub-states may exist in any real device~\cite{ArmLimited2021}.

% GOOD introduction about https://arxiv.org/pdf/1706.05715.pdf

\section{Related work}\label{related-work}

Over the years, several works have closely aligned with our approach. One of the earliest contributions to runtime attestation on lower-end processors was made by \textit{Hristozov et al.}\cite{hristozov2018practical}, who developed a runtime attestation scheme based solely on an \gls{MPU}, aiming to create a processor-agnostic solution. However, this approach lacks continuous monitoring, offering only on-demand measurements during execution. \textit{Huo et al.}\cite{Huo2020} also leveraged the \gls{MPU} for isolation but incorporated \gls{CFG} to detect control-flow vulnerabilities. Other works, such as \textit{Wang et al.}~\cite{Wang2023}, focused on addressing real-time constraints while defending against both control-flow and data-only attacks.
% Tirar para cima se for para cortar

In the \gls{CFI} domain, researchers have also investigated the potential of \glspl{MPU} to develop processor-agnostic solutions~\cite{Almakhdhub2020, Zhou2020, Choi2024}. Additionally, works by \textit{Nyman et al.}\cite{Nyman2017}, \textit{Kawada et al.}~\cite{Kawada2020}, and \textit{Yeo et al.}~\cite{Yeo2022} have explored the use of ARM Trusted Zone for ARM Cortex-M, focusing on enhancing the isolation of their solutions. Our approach aligns with these efforts, utilizing technologies such as ARM TrustZone and \gls{TF-M} as the \gls{TCB}. However, by requiring a processor that supports \gls{PACBTI}, we introduce both constraints and advantages. While this requirement limits the range of applicable systems, it also enables us to reduce \gls{CFI} overhead and improve energy efficiency.

Researchers have actively pursued hardware-based solutions for \gls{CFI} to minimize or eliminate performance delays. Notable systems include LO-FAT~\cite{Dessouky2017} and LiteHAX~\cite{Dessouky2018}, both designed for the RISC-V architecture. LO-FAT introduces a hardware-based runtime attestation system that adds an \gls{IP} core to record control-flow data for unmodified applications with minimal overhead. LiteHAX, built upon a custom processor extension, further advances this approach by recording data-flow events in addition to control-flow events. This enables LiteHAX to effectively address both control-flow and non-control data vulnerabilities, providing a more comprehensive solution to system security.

For the ARM Cortex-A architecture, LAHEL~\cite{Arias2020} employs a similar \gls{IP} core approach but integrates with existing debug features to conduct control-flow attestation. In contrast, \textit{Geden et al.}~\cite{Geden2023} pursued a different approach by developing a hardware security module that connects to the system bus to perform control-flow attestation externally. While this method is less invasive, it requires modifying the processor to enable access to the bus.

In \gls{CFI}-specific research, similar efforts have been made to mitigate performance impacts. For example, \textit{Davi et al.}~\cite{Davi2015} introduced a hardware extension in a simulated processor to reduce the performance penalties associated with \gls{CFI}. Other studies have focused on ensuring the compatibility of \gls{CFI} with real-time systems, aiming to maintain strict timing requirements without compromising performance~\cite{Walls2019}.

In contrast to these solutions, \textit{RunPBA} utilizes an existing hardware extension rather than requiring custom hardware modifications or additional \gls{IP} cores. This approach allows \textit{RunPBA} to operate on off-the-shelf processors, which significantly enhances its practical applicability. \textit{RunPBA} avoids the complexities and costs associated with custom extensions, making it an accessible solution for implementing runtime protection across a broader range of devices without sacrificing security.

Currently, there is no research specifically leveraging the \gls{PACBTI} extension on ARM Cortex-M for runtime attestation or security systems. However, this extension is inspired by the \gls{PAC} and \gls{BTI} features developed for the ARM Cortex-A series. Research has already explored the application of \gls{PAC} to \gls{CFI}. For instance, \textit{Liljestrand et al.}~\cite{Liljestrand2019} were the first to create an LLVM compiler extension that utilizes \gls{PAC} to prevent runtime attacks. Building on their work, several studies have further investigated the potential of \gls{PAC} to strengthen \gls{CFI} mechanisms~\cite{liljestrand2020pacstackauthenticatedstack, nasahl2021protecting, fanti2022toward, Zhang2021a, denis2020camouflage}. Additionally, although still relatively recent, research on the \gls{BTI} extension has emerged. Notably, \textit{Ammar et al.}~\cite{ammar2024bridging} introduced CFA+, a practical hardware-assisted control-flow attestation mechanism that leverages \gls{BTI} on the Cortex-A architecture.

In summary, extensive research has explored the use of \gls{CFI} techniques for enabling runtime attestation, with particular emphasis on adapting these methods for lower-end processors, such as the ARM Cortex-M series. These approaches typically rely on either software-based \gls{CFI}, which can incur significant overhead, or custom hardware modifications to mitigate performance impacts. The \gls{PACBTI} extension on Cortex-M introduces new opportunities in this research field, offering the potential to reduce the execution delays traditionally associated with \gls{CFI} without the need for custom hardware. \textit{RunPBA} is the first to leverage \gls{PACBTI} in this context, uniquely positioning this work within runtime attestation research by offering a solution that bypasses the limitations of software-only approaches and avoids the costs associated with specialized hardware changes.

%\textbf{Performance} and instrumentation??, energy

%PACStack: an Authenticated Call Stack

% https://www.ndss-symposium.org/wp-content/uploads/2018/07/diss2018_11_Hristozov_paper.pdf se for necessário melhorar o related work

% Talk about CFI SuM: Efficient shadow stack protection on ARM Cortex-M que usa hardware para proteger os registos

%\section{PACBTI}\label{pacbti}
\section{PACBTI Security risks}\label{pacbti-security-risks}

\acrfull{PACBTI} extension is an optional \gls{ISA} extension introduced with the ARMv8.1 revision for Cortex-M, and consists of two primary components: \acrfull{PA} and \acrfull{BTI}~\cite{Mujumdar2021}. Both were inspired by existing \gls{ISA} extensions in the Cortex-A architecture, carrying over many of their features and characteristics. This technology aims to protect devices from \gls{ROP}~\cite{Shacham2007} and \gls{JOP}~\cite{Bletsch2011} attacks by verifying code integrity and controlling the execution flow. As of now, the only two processors implement this extension, Cortex-M85 and Cortex-M52.

Naturally, like any other technology, certain risks have already been identified that could compromise \gls{PA} and \gls{BTI}.  Some of these risks target the underlying concepts themselves, while others focus specifically on the implementation details. In this subsection, we will explore these security risks in more depth. Many of them were initially aimed at the Cortex-A implementation, and we will examine if they are applicable to the Cortex-M. Previous research has highlighted four main types of attacks against \gls{PA}: brute-force attacks, malicious \gls{PAC} generation, reuse attacks, and side-channel attacks~\cite{Liljestrand2019, Brand2019, Ravichandran2022}. Against \gls{BTI}, one attack was discovered, \gls{FOP}~\cite{Guo2018,Stratton2023}.

\textbf{Brute-force} - As \gls{PAC}s are stored in memory or registers without any special security measures, they can be changed by an attacker during runtime. Despite that, the \gls{PAC} verification should fail if the pointer does not correspond to a valid signature. Some researchers have pointed out that, depending on the size of the \gls{PAC}, an attacker could leverage this fact to brute-force a correct  \gls{PAC} for an arbitrary address~\cite{Liljestrand2019, Brand2019, Ravichandran2022}. This risk is amplified by the fact that in \textit{Cortex-A} architectures, the \gls{PAC} is stored in the remaining bits of the address space, and its size varies based on the address scheme is use\footnote{Typically, Linux systems with \gls{PA} enabled use a PAC between 16 and 24 bits~\cite{Liljestrand2019}.}, ranging from 3 to 31 bits~\cite{QualcommTechnologies2017}. Therefore, the effectiveness of this attack will depend on the target configuration. In addition, programs typically use a random \gls{PA} key for each execution. Thus, any unsuccessful brute-force attempt will cause a crash, making the key change. \textit{Liljestrand et al.}~\cite{Liljestrand2019} suggests that an attacker could target a sibling of a pre-forked or multi-threaded program  to bypass this issue since it shares the \gls{PA} key with its parent process, meaning that a \gls{PAC} failure in a sibling process does not result in the termination of the parent process.

The \gls{PA} implementation on Cortex-M is not immune to brute-force attempts. However, because the \gls{PAC} value is stored in a separate register and is always a 32-bit value, a brute-force attack on Cortex-M is more challenging compared to the Cortex-A \gls{CPU}. However, the heterogeneity of embedded systems may make brute-force attacks easier, particularly if poor key management practices are used. To mitigate this risk, systems that use \gls{PA} in Cortex-M should ensure the use of adequate entropy and freshness of the \gls{PA} keys whenever a key change is required. Long-lived keys should be avoided to further reduce the potential for successful attacks.

% entropy of pac
% freshness of the key
% handling of the key on fail verifictions
% lack of aslr ??

\textbf{Malicious \gls{PAC} generation} - \gls{PA} is only a way to prevent a specific attack. \gls{PAC} generation occurs dynamically during the execution of the software, which means that the instructions to generate \gls{PAC}s are embedded in the software. An attacker could exploit this fact to generate arbitrary \glspl{PAC}, if they are able to control either the pointer being authenticated or the modifier.  However, this assumes that the attacker has already gained control of the program execution by other means.

\textbf{Reuse attacks} - Another potential way to compromise \gls{PA} is through a simpler method:  an attacker might be able to read a \gls{PAC} and latter reuse it. The purpose of the modifier in \gls{PA} is to prevent this. However, as mentioned earlier, the stack pointer is commonly used as the \gls{PA} modifier, which is not necessarily unique to a specific function invocation. This can be exploited by attackers to reuse authenticated pointers in different function calls, either by targeting a function with the same stack pointer or by forcing a function to have the same one.  Moreover, since the Cortex-M implementation has fewer keys that are used for longer periods of time, \glspl{PAC} are more susceptible to reuse attacks.

% recursion of the PA may lead to attack
% In Cortex-A the use of multiple keys is a way to mitigate this, therfeore have a single key for a secuity context doesn't help.

\textbf{Side-channel attacks} - \textit{Ravichandran et al.}~\cite{Ravichandran2022} discovered a speculative execution attack on Cortex-A processors that can be used to brute-force \glspl{PAC}. As explained earlier, Cortex-A \gls{PA} does not raise an exception when \gls{PAC} verification fails; instead, it makes the pointer invalid and expects an illegal address exception to be generated when the pointer is used. Attackers may exploit this behavior by using the branch prediction feature to deceive the processor into executing a verification operation during speculative execution. The result of this operation is leaked through a microarchitectural side-channel attack without causing any crashes and, consequently, key resets. 

This vulnerability does not affect the Cortex-M implementation, as the verification instruction generates a synchronous fault~\cite{ArmLimited2023a}. However, it raises the question whether other side-channel attacks could affect this implementation. Generally, low-end processors are not susceptible to such attacks because they lack speculative execution. Although classified as a low-end processor, the Cortex-M85, which supports \gls{PACBTI}, incorporates a branch prediction mechanism~\cite{ArmLimited2020}. This feature has been the source of several vulnerabilities in other architectures. Consequently, further research is needed to determine whether it introduces any security risks to \gls{PA}, as this type of technology is uncommon in processors within this range.

\textbf{\acrfull{FOP}} - Recently, \gls{FOP} has emerged as a new class of code reuse attacks. \gls{FOP} uses a \textit{dispatcher gadget}, which is basically a controllable indirect branch within a loop, as a way to execute arbitrary functions. This technique can successfully bypass coarse-grained \gls{CFI} and some fine-grained \gls{CFI} technologies~\cite{Guo2018}. Unfortunately, \gls{BTI} is also vulnerable to this class of attacks. \gls{BTI} enforces that any indirect branch jumps to a \textit{landing pad} instruction, which includes a \gls{PA} generation instruction. These instructions are located at the beginning of every function when \gls{PA} instructions are enabled during compilation. Therefore, \gls{BTI} does not restrict this technique in any way, as the attacker can jump to any function since all functions are valid \textit{landing pads}~\cite{Stratton2023}. Moreover, even when \gls{PA} is disable and only \gls{BTI} is enabled, several compilers automatically add \gls{BTI} landing pads to the beginning of functions. This behavior was observed during our researched. More details are provided in Appendix \ref{append:btifeature}.

Research on this attack has been limited and most has focused on the \textit{x86} architecture~\cite{Guo2018}, and existing research on Arm only targets Cortex-A~\cite{Stratton2023}. Additionally, unlike other code reuse attacks, for which numerous tools are available to identify gadgets and even create exploits, this is not the case for \gls{FOP}. To our knowledge, there is a single tool available online to automatically identify \gls{FOP} gadgets, \textit{FOP\_Mythoclast}~\cite{LMS572023}, that works with \textit{ARM Cortex-A} and \textit{x86} core files. \textit{Stratton et al.}~\cite{Stratton2023}, the team responsible for the development of this tool discussed the challenges involved in creating such tools and, ultimately, why these attacks are difficult to exploit in practice. The main challenge lies in identifying \textit{dispatcher gadgets}: in addition to finding a code segment that has a loop and an indirect branch, it is necessary to determine whether it is possible to control the target of the indirect branch, control the loop's iteration count, and avoid interference with registers crucial for executing \textit{function gadgets}. 

Therefore, despite \gls{FOP} being an attack that can compromise the most effective security measurements, it is difficult to put into practice. This technique requires extensive expertise and significant time investment from the attacker to create an exploit, as exploit chains need to be manually crafted. Researchers have pointed out that mitigating this risk requires the development of hardened compilers that avoid creating gadgets or zeroing out parameter registers at the end of each function~\cite{Stratton2023}. Regardless, there is still a lack of research on this attack targeting the Cortex-M processor family, making it impossible to fully understand its potential consequences on this processor family and the availability of gadgets in embedded systems.\\ 

%AVAILABILITY? and prior vulnerability

%Finally, it is noteworthy to mention that, similar to the other attacks we pointed out throughout this section, this attack requires a previous vulnerability that can lead the attacker to control the dispatch gadget.

% talk about the different registers and r12 register?

In summary, there are several risks that can jeopardize \gls{PACBTI}. However, every attack relies on a prerequisite primitive within the system, such as the ability to perform arbitrary reads and writes. Even when these conditions are met, the attacks are far from straightforward. In the case of \gls{PA} attacks, attackers must gain control over multiple aspects of the \gls{PA} generation process and ensure that the program is in a state that allows repeated execution of instructions at will. For instance, in reuse attacks, it is necessary to control the modifier either by directly controlling it or influencing it to be the same. To generate malicious \gls{PAC}s, the attacker must control the arguments involved in the \gls{PAC} generation. Brute-force and side-channel attacks require even more control; the program must be instrumented to systematically repeat the same set of instructions. Furthermore, to succeed, the attacker will need to repeatedly target multiple \gls{PAC}s in the same execution to be able to create a \gls{ROP} chain. Likewise, performing a \gls{FOP} attack against \gls{BTI} requires the presence of gadgets suitable for this technique, along with very specific conditions for program execution control. 

Finally, it is important to emphasize that technologies like \gls{PACBTI} should not be used in isolation. The incorporation of additional protections, such as \gls{ASLR} and \gls{NX} bits, can further hinder exploitation attempts targeting \gls{PACBTI}. As demonstrated, most attacks against \gls{PACBTI} require extensive access. Therefore, when combined with other security measures, the difficulty of executing these attacks increases. However, it is also well-established that implementing many of these protections in embedded systems presents its own set of challenges~\cite{shao2022faslr}.

\section{RunPBA}\label{runpba}
Over the years, the demand for trustworthiness in embedded systems has increased, and with that, the development of attestation mechanisms has risen. Runtime attestation solutions have emerged as one of the most effective ways to establish trust in remote devices. Many of these solutions leverage \gls{CFI} to implement runtime attestation. \gls{CFI} is the most effective way to mitigate control-flow vulnerabilities. The idea behind this technique is to restrict the program to a set of predefined legitimate paths, represented by a \gls{CFG}.  This is achieved through runtime checks that enforce these policies and, therefore, making it impossible for attackers to jump to unintended places in memory.  Unfortunately, because of these checks, it is known that these runtime attestation solutions have an impact on the performance and memory use of embedded devices~\cite{Dang2015}. To address this, researchers have developed solutions with hardware acceleration. Most of these solutions use custom hardware, making them difficult to adopt due to their inherent cost and complexity~\cite{Dang2015}. To help address this issue, we created \textit{Runtime PACBTI Attesation} (RunPBA), an hardware-based runtime attestation that uses off-the-shelf processor features to reduce overhead. 

\subsection{Goal and Requirements}
From our analysis of the current state of runtime attestations and its challenges, we identified the following requirements that our solution must meet in order to contribute significantly to this research field.

\textbf{R1 - Performance} - \textit{Minimize the overhead induced by the runtime attestation feature.}  If the attestation system introduces significant overhead, such as consuming excessive computational resources, memory, or power, it can degrade the system's performance, potentially impacting its ability to perform its primary functions efficiently~\cite{Dang2015}.
    
\textbf{R2 - Commodity hardware} - \textit{Rely on existing features of off-the-shelf processors.} Many past approaches to this challenge relied on custom hardware, which increases the cost of implementing this system, hampering widespread utilization. Commodity hardware makes the system compatible with a broad range of devices~\cite{Davi2009, hristozov2018practical}.

\textbf{R3 - Existing software} - \textit{Build on top of an existing attestation scheme widely used in the industry.} Following this path facilitates the adoption of this solution in the real world. It eases the learning curve for developers that need to integrate this system,  interoperability with existing software ecosystems, and increases overall reliability since we do not need to reinvent core functionality that could introduce fundamental flaws. In addition, established systems have already been scrutinized and proven in real-world scenarios~\cite{Oh2017}.
    
\textbf{R4 - Fault traceability} - \textit{Enable tracing back to the origin of the runtime attestation failure.} This way the system can be used as an incident response tool, to deterministically pinpoint the origin of a vulnerability.

\subsection{Attacker capabilities}\label{attacker_capabilities}
During the development of RunPBA, our focus will be on an attacker capable of performing arbitrary read and write operations within the \gls{NSPE} address space, but with respect to the memory permissions (which include limitations introduced by \gls{DEP}, and other limitations depending on the privilege level and current \gls{MPU} configuration). Although canaries may be enabled, an attacker with these capabilities could easily bypass these protections. The \gls{NSPE} will not have any other \gls{CFI} solution enabled, aside from \gls{PACBTI}, which will be configured on both privilege levels of the \gls{NSPE}.

The capabilities we have defined are typically associated with scenarios in which attackers exploit software vulnerabilities that enable unrestricted access to memory (such as buffer-overflows or format string vulnerabilities). Our primary focus is on the \gls{NSPE}, due to its larger attack surface and its role in hosting the user application. In contrast, inside \gls{SPE} we have \gls{TF-M}, which has a smaller footprint, with a well-defined \gls{API} and a life cycle designed to minimize its exposure to potential attacks. Our model considers \gls{TF-M} as \gls{TCB}, and we assume that \gls{TF-M} is correctly configured and that its threat model is followed~\cite{TMP_threat_model}.  Therefore, an attacker even with unrestricted access to the \gls{NSPE} cannot compromise or interact in unexpected ways with \gls{TF-M} itself.

Using access to the \gls{NSPE} address space, an attacker may attempt to execute code in this environment by exploiting software vulnerabilities. These vulnerabilities may be present in either the priviliged or the unpriviliged environment of the \gls{NSPE}. During vulnerability exploration, attackers may attempt to brute-force  or reuse \gls{PAC}s and even create \gls{FOP} payloads. These attacks were previously discussed in Subsection \ref{pacbti-security-risks}. We assume that the attacker cannot generate a malicious PAC since it must first gain execution control in \gls{NSPE}. We also consider noncontrol data attacks, side-channel attacks, and physical attacks to be out of the scope of this threat model. Our primary focus is software vulnerabilities that could compromise the device, as these are the types of attacks that RunPBA aims to mitigate.

In short, we consider an attacker that has a read and write primitive in the \gls{NSPE}, regardless of the privilege level. The attacker can perform any software-based attack against \gls{NSPE} or any exposed component of the \textit{RunPBA}. Non-control data attacks, side-channels attacks, physical attacks, and attacks against the \gls{TF-M}. 

\subsection{Design}

At the core of RunPBA is the \gls{PACBTI} extension, which we leverage to address two common challenges in attestation solutions. First, \gls{CFI}-based runtime attestation approaches often introduce significant delays in code execution, prompting researchers to explore hardware solutions to mitigate these overheads. As we demonstrate, \gls{PACBTI} can be used to reduce these performance penalties (\textbf{R1}). Second, many hardware-based solutions are custom-built, increasing manufacturing costs and limiting their adoption. Since \gls{PACBTI} is a standard ARM extension already available in off-the-shelf processors such as the Cortex-M85, and Cortex-M52, RunPBA can operate on commodity hardware, with \gls{PACBTI} being its only hardware requirement (\textbf{R2}).

In summary, on the hardware side, enabling control-flow attestation with the \gls{PACBTI} extension involves configuring and enabling the extension. If the \gls{PACBTI} extension triggers a fault, it indicates an attempt to subvert the device's execution, and consequently, an attestation failure must be reported. This task is handled by the attestation system. Building a complete attestation system with both static and runtime attestation from scratch may not be the optimal approach, as it would add complexity to the already heterogeneous landscape and hindering real-world adoption. Therefore, \textit{RunPBA} is designed on top of \gls{TF-M}, expanding its existing attestation system (\textbf{R3}).

\begin{figure}[h!]
    \centering
    \includegraphics[width=0.50\textwidth]{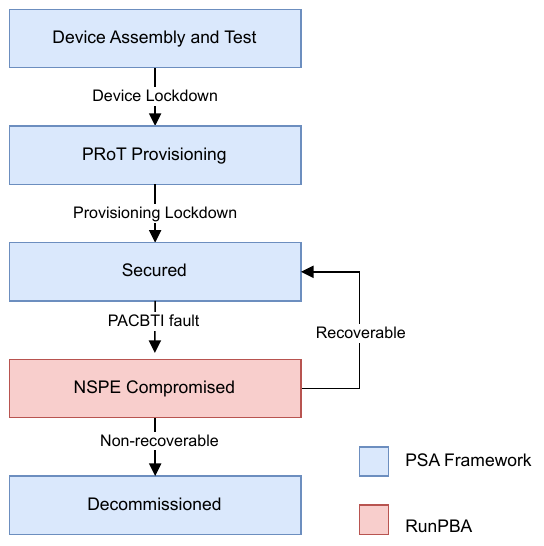}
    \caption{\gls{PSA} security lifecycle with \textit{RunPBA}}
    \label{psa_states}
\end{figure}

As mentioned before, the \gls{PSA} Security Framework envisions a device lifecycle, which is present in every attestation operation. With \textit{RunPBA}, we introduced a new state in the device lifecycle, called \textit{\gls{NSPE} Compromised}. A device enters this state whenever the \gls{PACBTI} extension causes a fault, indicating that a vulnerability in the \gls{NSPE} is being actively exploited. In this state, the control is not returned to the \gls{NSPE} until the situation is assessed (Figure~\ref{psa_states}). Depending on the assessment, the device can return to a secure state if recovery and vulnerability fixes are possible, or it can be decommissioned if security cannot be restored.

The strategy of holding the device in the \gls{SPE} until recovery may be disruptive to its utilization, or even exploited by an attacker to create a denial attack. However, this approach helps prevent further success by the attacker and mitigates potential brute-force attempts against \gls{PAC} (Subsection \ref{pacbti-security-risks}).

The \textit{\gls{NSPE} Compromised} state is further enhanced by the ability to trace the source of a \gls{PACBTI} fault. When \textit{RunPBA} detects a \gls{PACBTI} fault, it generates an object containing critical debug information that helps pinpoint the fault’s origin, facilitating device recovery after an attack (\textbf{R4}). This tracing capability leverages the isolation features of \gls{TF-M}, which enforces the separation between \gls{NSPE} and \gls{SPE}. \textit{RunPBA} uses this separation to mitigate attacks on the \gls{NSPE}, securely storing fault data within \gls{SPE} to ensure the continuity of essential services. In addition, this mechanism can integrate with an external incident response system to address the issue on the compromised device and issue a firmware update to resolve vulnerabilities on similar devices.

\subsection{Implementation} \label{runpba_implementation}

The \textit{RunPBA} system is implemented as an \textit{Application RoT} of the \gls{TF-M} environment, requiring only minimal modifications to the original \gls{TF-M} source-code, as we will explain in detail throughout this section. This system architecture ensures easy portability, as many hardware manufacturers maintain their own \gls{TF-M} forks, and improves the overall security level of the system by inheriting the security constraints associated with this type of component. 

The \textit{RunPBA}'s \textit{Application RoT} implements all the core features of the system, handling \gls{PACBTI} security faults, managing security fault records, and verifying the \gls{PACBTI} status within the \gls{NSPE}. These features are accessible to other \gls{SPE} components through a secure \gls{API}, while the \gls{NSPE} is unable to interact directly with the \textit{RUNPBA} system and only sees its results in the attestation result.  

The \gls{TF-M} implements the Attestation Service as a Platform \gls{RoT} partition, enabling applications to prove the device's identity, status, and configuration to a third party. It can be configured to use either an asymmetric or symmetric key. In both scenarios, each device is assigned a unique key for identification and to authenticate the attestation token, which is generated upon request. The output of this service is an attestation token which is composed by several claims with the information the device wants to attest. Among these claims, there is  the \textit{security lifecycle} claim, attesting the current state of the device~\cite{ArmLimited2024}. \textit{RunPBA} uses this claim to represent the \textit{\gls{NSPE} Compromised} state.

The \gls{PSA} framework allows custom claims in the attestation token. However, currently, the attestation partition does not provide a way to customize this token. Therefore, the upstream \gls{TF-M} implementation was modified to generate an attestation token that incorporates the output from the \textit{RunPBA} system. Another modification was necessary within the \gls{TF-M} core itself, as the default behavior upon an exception is to reset the device. Therefore, to support \textit{RunPBA}, it was essential to handle specific exceptions triggered by \gls{PACBTI}.

\subsubsection{Fault handling and traceability} 
In the Cortex-M architecture, exceptions are not agnostic to the security levels introduced by the Arm TrustZone extension. Many faults are handled independently based on the environment in which they are generated, which means that the device maintains a distinct set of handlers and configurations for each security level. One of the faults that can be banked between the two states is the \textit{Usage Fault} exception. This particular exception is triggered when a \gls{PACBTI} verification fails. Thus, if a failure occurs within the \gls{NSPE}, it raises a \textit{Usage Fault} exception in the non-secure world.

When a \gls{PACBTI} verification failure occurs, we aim to stop execution in  the \gls{NSPE} entirely due to the potential severity of the attack. To achieve this, \textit{RunPBA} requires the \textit{Usage Fault} system handler to be disabled (with \textit{SHCSR\_NS.USGFAULTACT} set to 0) and all \textit{Hard Faults} to be redirected to the secure state (by setting the \textit{AIRCR.BFHFNMINS} bit to 1). With this setup, when a \textit{Usage Fault} is triggered in the \gls{NSPE} without an active system handler, the processor escalates the fault to a \textit{Hard Fault}. Since \textit{Hard Faults} are the highest priority exceptions, this configuration ensures that the fault is handled immediately and within the secure state.

\textit{Hard Faults} can be caused by various issues in a device, not only \textit{Usage Fault} triggered by the \gls{PACBTI}. For instance, \textit{Hard Faults} can arise when exceptions are triggered during other interrupts. Therefore, before \textit{RunPBA} is triggered, the source of the fault must be accurately identified. 

The Configurable Fault Status Register(CFSR) can be used to indicate the cause of the current fault. In the case of a \gls{PACBTI} fault, the non-secure CFSR should indicate a \textit{Usage Fault} caused by an \textit{Invalid State}. However, an \textit{Invalid State} can still have other origins, such as a simple division by zero. The current design does not provide a direct mechanism for uniquely identifying \gls{PACBTI} faults. This limitation means that further debugging is necessary to differentiate \gls{PACBTI} faults from other Invalid State errors.

As discussed previously, \gls{PACBTI} is a combination of two mechanisms: \gls{PAC} and \gls{BTI}. If a fault occurs during \gls{PAC} verification, the stored Program Counter~(PC) will point to a verification instruction. In the case of a \gls{BTI} failure, the \textit{EPSR.B} will be set, indicating a branch to non-landing pad instruction. \textit{RunPBA} uses these two indicators to distinguish \gls{PACBTI} faults from other types errors. This is feasible because, during a context switch, the device's state is saved in memory to enable resumption of execution. \textit{RunPBA} uses this saved state to establish the cause of the fault and enhance fault traceability.

Identification of \gls{PACBTI} faults can be performed within the fault handler. However, to securely store this information, it must be communicated to the \textit{Application RoT}. Partitions in \gls{TF-M} expose methods for interaction with other \gls{TF-M}'s components, but these methods are not accessible during fault handling. Typically, these methods operate through function calls in which parameters and call information are stored in a message system, and an interrupt is triggered to dispatch the partition thread to handle each request. These calls expect to run in thread mode, whereas an exception triggers the processor to execute in handler mode. Additionally, a \textit{Hard Fault} runs at the highest priority level to avoid being interrupted. 

An alternative way for interacting with an Application RoT is through interrupts. Partitions can define two types of handlers: \gls{FLIH} and \gls{SLIH}. \gls{FLIH} runs in handler mode and is triggered immediately, but has limited \gls{API} access due to its special exception context. On the other hand, in \gls{SLIH}, execution is deferred in time and runs in the secure partition thread, allowing full access to the secure partition \gls{API}s. \textit{RunPBA} exploits these mechanisms to enable interaction with the \textit{RunPBA} from within the fault handler.

The \textit{RunPBA} partition includes two handlers: a \gls{FLIH} that receives the fault context as an argument and stores a copy in the \textit{Application RoT} memory, and a \gls{SLIH}, which is linked to an unused physical GPIO interrupt. This handler is responsible for transferring the fault context from the \textit{Application RoT} buffer to the \gls{ITS}. In more detail, when a \textit{Hard Fault} handler detects a PACBTI-triggered exception, it invokes a specific \gls{FLIH} in the \textit{RunPBA} partition though the \gls{SPM}. This \gls{FLIH} copies the fault context to a buffer within the \textit{Application RoT} memory. After completing this call, the \textit{Hard Fault} handler triggers the \textit{RunPBA} partition once again, this time through the GPIO interrupt. This action prompts \gls{SPM} to schedule the \textit{RunPBA} thread in \gls{SLIH} mode, ensuring that the fault and its context are properly stored in the \gls{ITS}. The source address of each fault can subsequently be retrieved for future analysis via an \gls{API} made available to other \gls{SPE} applications.

\begin{figure}[ht!]
    \centering
    \includegraphics[width=0.45\textwidth]{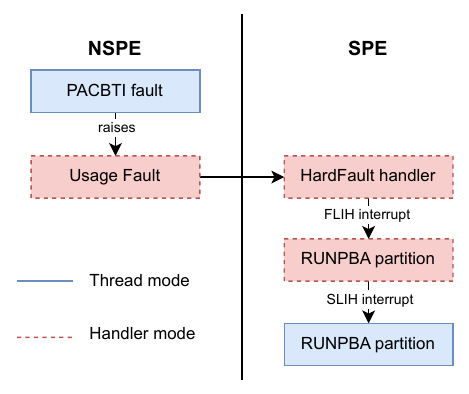}
    \caption{\gls{PACBTI} fault handling with \textit{RunPBA}}
    \label{runpba_fault_handling}
\end{figure}

Between the \gls{FLIH} and \gls{SLIH} processes, the fault handler also redirects the \gls{NSPE} execution address to an infinite loop. This ensures that the \gls{NSPE} does not resume execution since, for the \gls{SLIH} interrupt to be processed, the program must exit handler mode and enter thread mode. Although \gls{SPE} takes priority over the \gls{NSPE} for scheduling, this step further guarantees that \gls{NSPE} remains stopped. Following this setup, the \textit{RunPBA} application manages the subsequent execution of the device, which may involve resetting the device or putting it on hold until the fault is analyzed. All this process is illustrated in the Figure~\ref{runpba_fault_handling}.

\subsubsection{Attestation}

The \gls{PSA} Initial Attestation Service is the \gls{TFM} application responsible for attesting the device's identity. This service generates a signed token following the Arm's \gls{PSA} Attestation~\cite{Tschofenig2024} model. In this model, the verifier challenges the attestation service with a nonce that must be contained in the response to verify its freshness. The \gls{PSA} Initial Attestation Service uses a private key that only the device possesses to sign a token containing several claims that express the current state of the device and represent other device-specific data. For instance, the nonce provided by the verfier is included in the token as a claim. Some claims are mandatory, such as the security lifecycle claim, which indicates the current lifecycle state of the device, while others, such as the hardware version claim, are optional~\cite{ArmLimited2022}. From an implementation point of view, the token is formatted as a CBOR Web Token and adheres to the IETF Entity Attestation Token standard~\cite{Lundblade2024}.

The \textit{RunPBA} builds on top of this service to represent its state during the attestation process. Instead of introducing a new claim in the existent attestation token, \textit{RunPBA}  expands the security lifecycle claim. The security lifecycle claim is represented as a 16-bit array, where the final 8 bits denote the Platform \gls{RoT} security lifecycle state and the first 8 bits are implementation defined. \textit{RunPBA} utilizes these bits to represent its state. This configuration ensures that the \textit{RunPBA} state is always present in any attestation operation, providing the relying party with real-time insight into the system's status.

\begin{figure}[h!]
    \centering
    \includegraphics[width=0.50\textwidth]{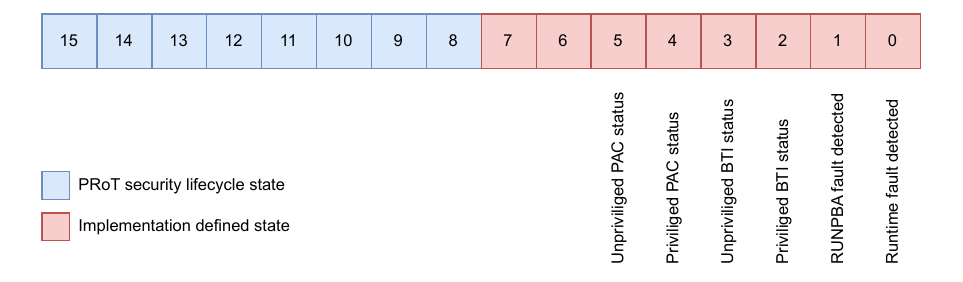}
    \caption{Representation of the Security Lifecycle claim with the \textit{RunPBA} system in the attestation token}
    \label{fig:lifecyclecClaim}
\end{figure}

The first bit of the security lifecycle claim indicates whether a runtime failure has occurred, signaling that the device is in the \gls{NSPE} compromised state. The second bit is set to 1 if a malfunction occurs within the \textit{RunPBA} system itself, such as when the \textit{RunPBA} application is unable to access secure storage. The remaining bits represent the status of each \gls{PACBTI} feature: a value of 0 indicates that a feature is disabled, while a value of 1 signifies that the feature is enabled (Figure~\ref{fig:lifecyclecClaim})

These details are collected by the initial attestation service during token generation. The service queries the \textit{RunPBA} application via \gls{PSA} \gls{API}. In response, the \textit{RunPBA} application checks for any \gls{PACBTI} faults and reads the control registers of \gls{PACBTI} to verify which features are enabled. This is possible because applications in the \gls{SPE} have access to the \gls{NSPE} memory, which allows them to gather the necessary data for an accurate attestation.

\subsection{Detailed overview}
\begin{figure*}[h!]
    \centering
    \includegraphics[width=0.92\textwidth]{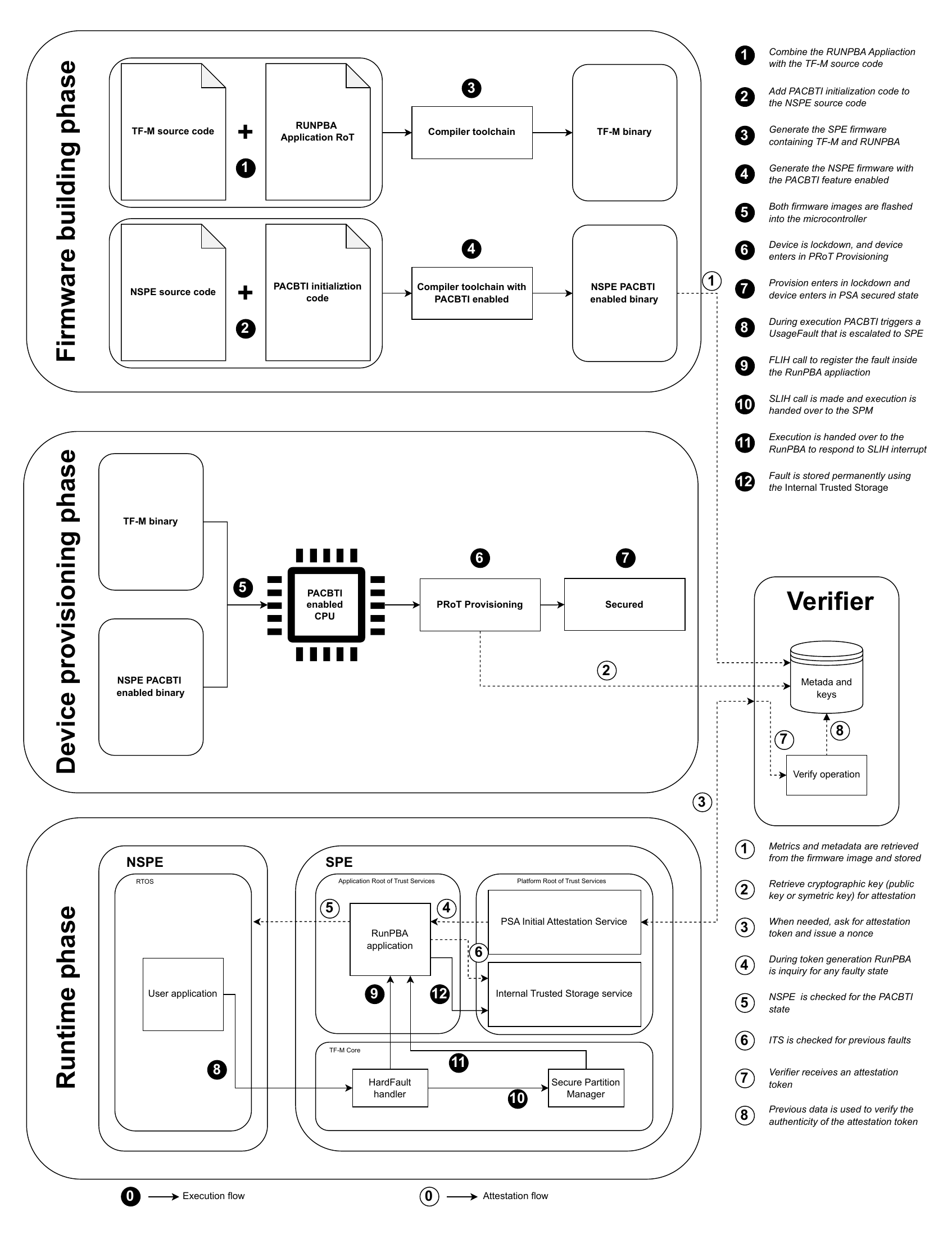}
    \caption{\textit{RunPBA} lifecycle overview}
    \label{runpba_overview}
\end{figure*}

To better picture the overall design of the \textit{RunPBA} system, Figure \ref{runpba_overview} depicts an overview of the system design. During firmware development, both the TF-M and NSPE applications must be adapted.  \textit{RunPBA} Application needs to be added to the \gls{TF-M} source code \blackCircle{1}, and on the \gls{NSPE} side, the PACBTI should be properly initialized \blackCircle{2}, using a secure random number generator for the \gls{PACBTI} key on each boot, and an infinite loop should be added. This function may change according to the embedded \gls{OS} used, since some of them may already have built-in initialization procedures, while others must be created by the developer. Finally, in the case of the \gls{NSPE} code, the source code must be compiled with a compiler aware of \gls{PACBTI}, to add the necessary instructions to the firmware \blackCircle{4}.  No further changes are needed in the firmware building phase of the device. In this phase, it is also possible to retrieve firmware metrics for the attestation system \Circled{1}. This is not specific to the \textit{RunPBA} system, but rather to the \gls{TF-M} attestation procedure. 

\textit{RunPBA} does not introduce any changes in the device provision phase. Both firmware images are flashed into the microcontroller \blackCircle{5} and the device is locked down, no debug features are enabled. It is also during provisioning that the device generates its own cryptographic key \blackCircle{6} for attestation \Circled{2}, which is generated and stored inside the \gls{RoT}.

When the device is put into production, if a \gls{PACBTI} failure is detected, it will lead to an \textit{UsageFault} within the \gls{NSPE} \blackCircle{8}. This fault escalates to a \gls{SPE} hard fault. The \gls{SPE} hard fault handler is then responsible for communicating the fault to the \textit{RunPBA} application using interrupts \blackCircle{9}, and scheduling the execution of \textit{RunPBA} \blackCircle{10} \blackCircle{11}. The \textit{RunPBA}
application will then store the fault and its origin in a secure way \blackCircle{12}.

On the attestation side, the only difference compared to the standard TF-M attestation system is that during the generation of the attestation token, the attestation service queries the \textit{RunPBA} application \Circled{4} for past \gls{PACBTI} faults \Circled{6} and the current status of the \gls{PACBTI} in the \gls{NSPE}  \Circled{5}.

\subsection{Limitations}\label{limititations}

The majority of our implementation limitations are the result of the inherent characteristics of \gls{PACBTI}. As mentioned in Subsection~\ref{pacbti-security-risks}, \gls{PACBTI} is vulnerable to \gls{FOP} attacks because \gls{BTI} instructions allow any function to be used as a landing pad, not limiting it to the intended one. Consequently, \textit{RunPBA} inherits this problem. At this point, the most viable solution to mitigate this issue is the use of hardened compilers that are specifically designed to prevent the generation of dispatcher gadgets, which could be exploited in \gls{FOP} attacks.

Moreover, by attacking a device with \gls{FOP}, the \textit{RunPBA} system could still be vulnerable to the \gls{TOCTOU} problem. An attacker exploiting \gls{FOP} might temporarily disable \gls{PACBTI}, if the appropriate gadget is available, creating an opportunity for further exploitation. If the attacker re-enables \gls{PACBTI} before the next attestation request, \textit{RunPBA} may fail to detect that its security mechanism has been compromised.

However, this type of attack would be highly complex and require two specific conditions that are typically difficult to satisfy. First, a complete gadget chain with the correct instruction to disable \gls{PACBTI}. Second, the attack must be executed in a privileged environment, which is uncommon, as most software exposed to attacks typically operates in an unprivileged mode. Furthermore, most software is designed only to enable \gls{PACBTI}, not disable it, rendering a viable chain of instructions to disable \gls{PACBTI} virtually nonexistent, since the sequence of instructions are not present in the firmware. While this attack vector is theoretically possible, it remains unlikely in typical real-world scenarios.

Moreover, during the \textit{RunPBA} development, we investigated the possibility of proactively monitoring any change in the \gls{PACBTI} status. Among the different alternatives, the most promising one was the ARM Data Watchpoint and Trace unit, as it has been previously utilized in this field by other solutions. Unfortunately, the registers responsible for controlling this feature cannot be monitored, since they are part of the processor's core registers and lack a representation in the address space.

Therefore, under the current design of \gls{PACBTI}, it is not possible to introduce additional constraints to control access to \gls{PACBTI}. From the \gls{NSPE}, only software running in privileged mode can enable or disable \gls{PACBTI}. Although this condition reduces the risk of exploitation, it cannot completely eliminate the potential for compromise.

Finally, even though most of our code is self-contained in an Application \gls{RoT}, \textit{RunPBA} requires changes to the upstream version of \gls{TF-M}. As with any change, this introduces implications in terms of security and maintainability. Changes to the upstream version were kept to a minimum. In terms of attestation, we added code to retrieve the claim generated by the \textit{RunPBA} application, and append this information to the attestation token. This code is contained inside the attestation partition; therefore, any problem that may arrive from this change cannot affect other partitions. On the other hand, the change in the hard fault handler is done to the \gls{TF-M} core. The change in the attestation partition could be removed in a future iteration of this work, using a delegated attestation mechanism. However, the changes in the hard fault handler will always be needed to \textit{RunPBA} to work.

\subsection{Security analysis}

The attacker described in the Subsection~\ref{attacker_capabilities} is able to perform any read and write operation on the \gls{NSPE} memory space, which means that the attacker can attempt to control the program execution or even disrupt the \textit{RunPBA} system. In this section, we will perform a security analysis according to the attackers capabilities against the \textit{RunPBA} system.

Attackers cannot interact directly with \textit{RunPBA}, as it does not expose any interface to \gls{NSPE}. However, they may use processor interrupts, during the handle of the PACBTI fault, to stop the execution of the \textit{RunPBA} system, and cause the processor to return to the \gls{NSPE} execution. The optimal moment to execute this attack occurs when the system transitions from managing the \textit{RunPBA} fault to running the \textit{Application RoT}. If the attacker is able to generate an interrupt, the processor will first schedule every \gls{SPE} interrupt before executing the ones related to \gls{NSPE}. Additionally, even if the attacker can schedule a \gls{SPE} interrupt that leads to the execution of the \gls{NSPE} code, there is a small time window to do it,  since right before the \gls{PACBTI} fault handler handed over the execution,  it cleans every active interrupt and only leaves the one responsible for executing the \textit{RunPBA} application and the interrupt should have an higher priority than the \textit{RunPBA} interrupt. Moreover, \textit{RUNPBA} also overwrites the \gls{NSPE} execution address with an address of an infinite loop, during the fault handling, has a second measure to prevent execution of malicious code.     

As explained in Subsection~\ref{pacbti-security-risks}, there is a risk that the attacker will use brute force to bypass PACBTI. There are two measures that work together to mitigate this risk, one on the embedded \gls{OS} side and the other on the \textit{RunPBA}.  The embedded \gls{OS} side is responsible for following the ARM recommendations, using a random \gls{PAC} key on each device initialization, which should not be difficult to implement since the \gls{TF-M} offers, as a primitive, a secure random number generator. On the RunPBA side, if a \gls{PACBTI} operation fails to authenticate a tag, it activates the RunPBA system, which registers the failure and triggers a reset of the device. This means that any brute-force attempt would be interrupted since during boot the \gls{PAC} key should change. Moreover, even if there is a failure in the embedded \gls{OS} and the key does not change, on a subsequent attestation operation, the \gls{PACBTI} fault will be stated in the attestation token.

The attacker, if the necessary gadgets are available, can perform a \gls{FOP} attack against the \gls{NSPE} and not be detected by the  \textit{RunPBA}. In this case, an attacker would attempt to disable the \gls{PACBTI} system to be able to execute code without its restrictions. Consequently, the  \textit{RunPBA} will loose its visibility of the \gls{NSPE}. An attacker will only be able to perform this action if it is executing in privilege mode, which decreases the probability of this attack. Furthermore, during each attestation, \textit{RunPBA} verifies if \gls{PACBTI} is enabled on \gls{NSPE}, and therefore signals \textit{RunPBA} that something is happening if it is disabled. This fact is then stated on the attestation token. Nevertheless, there is still the possibility of a \gls{TOCTOU} vulnerability, since the attacker could disable and then enable \gls{PACBTI} between two attestation operations and would not be detected. During the development phase, we investigated several options to mitigate this risk, but without success, the investigation carried out was described in more detail in Subsection~\ref{limititations}.

% Proof of execution
% FOPs

% https://arxiv.org/pdf/1706.05715 ??

\section{Evaluation}\label{evaluation}
To evaluate the proposed solution, we conducted two types of tests: functional and performance. The functional evaluation aimed to verify whether the system behaved as expected, and was performed in a virtualized environment using the Arm Fixed Virtual Platform (FVP) for Arm Corstone-310, a reference design incorporating the Cortex-M85 processor. The performance evaluation, on the other hand, focused on measuring the computational overhead introduced by the solution and was conducted on a Renesas EK-RA8M1 development board, which features the Cortex-M85 processor. This dual approach provided a comprehensive assessment of both the system’s correctness and its operational efficiency.

Unfortunately, as of this writing, this board does not support the Trusted Firmware-M~\cite{Renesas_Trust_Firmware_2024,Renesas_Trusted_Firmware_Sample_2024}, limiting us to testing our implementation in a virtual environment. The complete source code for this software is available in the GitHub repository located at \url{https://anonymous.4open.science/r/runpba-26CD}.

\subsection{Functional}

Our \gls{PoC} implementation includes a small \gls{NSPE} application. This application first outputs the attestation token to the serial console, then provides an echo function that prints back anything written by the user in the serial console. Within this application, a secret is stored in memory, and there is two buffer-overflow vulnerabilities. Both very similar, the user message is copied into a buffer without checking if the message exceeds the buffer's size. However, one of them allows the attacker to target the forward-edge of the execution flow.

To demonstrate these vulnerabilities, we developed two exploit that can output the in-memory secret to the console. One attacking the authentication operation of the PAC (backward-edge protection) and the other attacking the BTI instruction (forward-edge protection). Both exploit leverage a \gls{ROP} chain to execute its payload, given that the device stack is non-executable. Additionally, after extracting the secret, the exploit restores the program to its normal execution flow, mimicking the behavior of an attacker trying to evade a \gls{TOCTOU} vulnerable attestation system.

Each exploit was executed against two versions of the program: one with \gls{PACBTI} (and thus \textit{RunPBA}) enabled and the other with \gls{PACBTI} disabled. When \gls{PACBTI} was disabled, the exploit was successfully executed, printing the secret message and then resuming the behavior of the intended program without interruption. In contrast, with \textit{RunPBA} enabled, the exploit triggered a fault, as expected. \textit{RunPBA} intercepted the fault, executed its pre-defined response actions, and then reset the device, effectively preventing the exploit from completing its attack. This behavior confirms that \textit{RunPBA} can detect and respond to security violations proactively, enhancing the device's resilience against buffer overflow and ROP-based attacks. Moreover, during this test, the attestation token was tested to ensure that it correctly displayed the state of the \textit{RunPBA} and \gls{PACBTI} status. 

\subsection{Performance}
 
For our performance evaluation, we were unable to directly test our Proof-of-Concept on the Renesas EK-RA8M1 due to its current lack of support for \gls{TF-M}. However, any performance overhead introduced by our system would closely mirror the impact of executing \gls{PACBTI}-instrumented code, as the only modification \textit{RunPBA} introduces to the \gls{NSPE} application is the inclusion of \gls{PACBTI} instructions. As a result, we concentrated on evaluating the performance impact of \gls{PACBTI} itself on a real processor.

In theory, \gls{PAC} could introduce a noticeable overhead due to the additional  instructions required per function call, as each function requires two new instructions for \gls{PAC}, increasing CPU cycles. In contrast, \gls{BTI} does not require new instructions; it is managed entirely by the hardware, which implies a lower performance impact. 

We evaluated two metrics: the time overhead introduced by \gls{PACBTI} and the increase in power consumption. For time overhead, we measured the time difference of running the same program with and without PACBTI instructions. For power consumption, we monitored energy consumption to assess additional demands from PACBTI.

Our evaluation relied on two established benchmarks: "BEEBS: Open Benchmarks for Energy Measurements on Embedded Platforms" developed by \textit{James et al.}~\cite{Pallister2013}, which has been widely adopted in academic research, and \textit{CoreMark-Pro}~\cite{EEMBC_CoreMark_PRO_2024}, a benchmark commonly employed in industry. These benchmarks allowed us to evaluate performance and energy impact in scenarios that reflect real-world applications.

For each benchmark test, \gls{PACBTI} was evaluated with both its enabled and disabled states to compare performance metrics. Execution time and power consumption were measured using the Nordic Power Profiler Kit II (PPK2)~\cite{Nordic_Power_Profiler_2024}, a high-precision power profiling tool capable of measuring currents as low as 500 nA at a sampling rate of 100,000 samples per second. The PPK2 was configured in \textit{Source Meter mode} to supply power to the Renesas EK-RA8M1 board. Its accompanying software recorded the test results, which were subsequently exported as CSV files for further analysis.

To monitor firmware image size, we used the Linux \textit{size} command to measure the \textit{text}, \textit{data} and \textit{bss} memory regions. A Python script was used to calculate the duration and current consumption of each test, along with the geometric mean. The benchmarks were compiled in the Renesas e² studio IDE (version 24.4.0)  using the LLVM Embedded Toolchain for Arm, version 17.0.1, with optimizations disabled (\textit{-O0}).  With optimizations enabled, the compiler would remove function calls, making code inline, and many times removing functions all together when its output was not being used. This would compromise any overhead measurement, since the number of \gls{PACBTI} instructions is minimized. The scripts, CSV files, source code, and IDE settings are available in the following GitHub repository: \url{https://github.com/MrSuicideParrot/pacbti-benchmark}.\\

\noindent \textbf{BEEBS} - The BEEBS benchmark suite ~\cite{Pallister2013} was selected for its architecture-independent focus on energy consumption in embedded processors, with tests designed to target specific CPU operations such as floating point arithmetic, memory access, and branching. This design provides a detailed view of how different types of operation impact the execution of the program. However, the suite's small tests often have very short execution times, a limitation noted in previous studies~\cite{Zhou2020}. To address this, each test was looped 1024 times, ensuring a sufficient execution duration for accurate measurements.

Of the initial 30 tests intended for evaluation, three (\textit{WIKISORT}, \textit{PICOJPEG}, and \textit{FRAK}) were excluded due to crashes caused by a stack alignment issue. Additionally, three tests (\textit{MATMULT\_INT}, \textit{SGLIB\_LISTSORT}, and \textit{SGLIB\_QUEUE}) produced unexpected results where execution times with PACBTI enabled were shorter than without it. The \textit{MATMULT\_INT} test showed the most noticeable effect, with a 8\% reduction in execution time compared to the baseline. This outcome occurred because the compiler, with \gls{PACBTI} enabled, consolidated all benchmark functions into a single function, leading to a more optimized execution. The other two tests exhibited smaller reductions of around 1\%, also due to minor changes in the assembly that made the execution slightly faster with \gls{PACBTI} enabled. Despite efforts to prevent this behavior, we were unsuccessful. Therefore, these results were excluded from geometric mean calculations to avoid artificially lowering the overhead and skewing the overall results (Table \ref{tab:pacbti-beebs-sum}).\\

\begin{table*}[h!]
\centering
 \resizebox{\textwidth}{!}{\begin{tabular}{l|ccc|ccc}
\hline
\multicolumn{1}{c|}{\multirow{2}{*}{\textbf{Metric}}} & \multicolumn{3}{c|}{\textbf{Baseline - Pacbti disabled}} & \multicolumn{3}{c}{\textbf{Pacbti enabled}} \\ \cline{2-7} 
\multicolumn{1}{c|}{}                                    & Time (ms)    & Current over time (A)    & Size (bytes)   & Time (X)  & Current over time (X) & Size(X) \\ \hline
\textbf{Min}     & 196      & 2,88E+09 & 5812  & 1,001 & 0,998 & 1,033                     \\
\textbf{Max}     & 126535,3 & 1,90E+12 & 22184 & 1,287 & 1,311 & 1,094                     \\
\textbf{Geomean} & N.A.     & N.A.     & N.A.  & 1,047 & 1,048 & 1,068 \\
\textbf{Median}  & N.A.     & N.A.     & N.A.  & 1,015 & 1,011 & 1,075 \\ 
\textbf{ 95\% Geomean CI}  & N.A.     & N.A.     & N.A. & [1,017 - 1,077] &  [1,013 - 1,082]  & [1,039 - 1,054] \\ \hline

\hline
\end{tabular}}
    \caption{\label{tab:pacbti-beebs-sum}Summary of the performance overhead on BEEBS}
\end{table*}

\noindent \textbf{CoreMark-Pro} - The  CoreMark-Pro~\cite{EEMBC_CoreMark_PRO_2024} is a widely used benchmark in the industry to evaluate all processors, including microcontrollers. Compared to the BEEBS benchmark , CoreMark-Pro offers more intensive tests, resulting in longer execution times and the ability to assess multicore processing. However, for our evaluation. we disabled the multicore tests as they are not relevant for our evaluation.  Additionally, CoreMark-Pro incorporates a built-in loop feature that allows each test to repeat a specified number of times, making it easier to obtain reliable measurements. We fine-tuned this loop count to gather consistent data across tests. 

Similarly to our findings with BEEBS, two CoreMark-Pro tests were executed faster with \gls{PACBTI} than when disabled. A review of the generated assembly code revealed similar optimizations, where the compiler, where \gls{PACBTI} enabled caused the compiler to restructure functions in a way that slightly improved execution time. Consequently, to avoid skewing results, we excluded any tests in which the \gls{PACBTI} enabled execution was faster than the baseline when calculating the geometric mean for this benchmark (Table \ref{tab:pacbti-coremark-summary}).

\begin{table*}[h!]
    \centering
    \resizebox{\textwidth}{!}{
\begin{tabular}{l|ccc|ccc}
\hline
\multicolumn{1}{c|}{}                                     & \multicolumn{3}{c|}{\textbf{Baseline - Pacbti disabled}} & \multicolumn{3}{c}{\textbf{Pacbti enabled}}     \\ \cline{2-7} 
\multicolumn{1}{c|}{\multirow{-2}{*}{\textbf{Metric}}} & Time (ms)    & Current over time (A)    & Size (bytes)   & Time (X)   & Current over time (X) & Size(X)    \\ \hline
\textbf{Min}     & 53,47    & 6,39E+08 & 29708 & 1,001 & 1,000 & 1,011 \\
\textbf{Max}     & 23614,23 & 3,67E+11 & 67444 & 1,026 & 1,029 & 1,070 \\
\textbf{Geomean} & N.A.     & N.A.     & N.A.  & 1,011 & 1,010 & 1,048 \\
\textbf{Median}  & N.A.     & N.A.     & N.A.  & 1,009 & 1,006 & 1,064 \\ 
\textbf{ 95\% Geomean CI}  & N.A.     & N.A.     & N.A. & [0,999 - 1,017] &  [0,999 - 1,019]  &  [0,996 - 1,094] \\ \hline

\end{tabular}}
    \caption{\label{tab:pacbti-coremark-summary} Summary of the performance overhead on Coremark PRO}
\end{table*}

\subsubsection{Result analysis}

Tables \ref{tab:pacbti-beebs-sum} and \ref{tab:pacbti-coremark-summary} summarize the overhead results of both BEEBS and CoreMark-Pro benchmarks when \gls{PACBTI} is enabled. Enabling PACBTI incurs an overhead of 1.1\% in execution time overhead with CoreMark-Pro and 4.7\% overhead with the BEEBS benchmark. The results of these benchmarks are available in our repository. 

In the BEEBS benchmark, the highest overhead,  28\%, was observed in the \textit{recursion} test, followed by the \textit{sqrt} test with 24\%. The \textit{recursion} test calculates the Fibonacci of 30 using recursion, resulting in frequent function calls. Each call incurs the additional execution of \gls{PACBTI} instructions, leading to a cumulative overhead. Similarly, the \textit{sqrt} calculates the square root of 100 numbers using the Taylor series, with repeated calls to a function responsible for modulus calculation within a loop. Transforming this modulus function into an inline function notably reduced the overhead to 11\%. Moreover, the median of the results was 1.5\%, substantially lower than the geometric mean. This indicates that most tests had a minor impact, but some high-overhead tests skewed the geometric mean.

For the CorMark-Pro benchmark, PACBTI had a lower overhead with a geometric mean of 1\% and a maximum overhead of  2.6\%, in the \textit{radix2-big-64k}  test. This \textit{radix2-big-64k} test performs a discrete Fourier transform on a sequence and involves numerous function calls, which aligns with the slight increase in overhead due to \gls{PACBTI}’s additional instructions.

The increase in energy consumption due to \gls{PACBTI} closely mirrors the observed execution time overhead. In the BEEBS benchmark, power consumption increased by 4. 7\%, while in the CoreMark-Pro benchmark, it increased by only 1\%, reinforcing the principle that computation time and energy consumption are generally proportional~\cite{pallister2015identifying}.

A potential concern was that \gls{PACBTI} operations, which rely on hardware-implemented cryptographic functions, could lead to a substantial increase in power consumption. This concern arises from the possibility that, while hardware optimizations might reduce execution time overhead, they could still result in significant energy demands. However, the results indicate that such impacts are minimal in practice.

The hardware supporting \gls{PACBTI} has the potential to substantially increase energy consumption due to its cryptographic operations, despite mitigating the time overhead of this feature. However, the results demonstrate that this concern is unfounded. The hardware implementation of \gls{PACBTI} does not lead to a significant increase in power consumption, confirming that enhanced security features can be adopted without introducing excessive energy demands.

Furthermore, enabling \gls{PACBTI} resulted in a minimal increase in firmware size, with an observed growth of up to 0.05\% in the BEEBS benchmark and 0.06\% in CoreMark-Pro. This increase is primarily influenced by the number of functions and indirect jumps within the firmware. Generally, the size increase correlates with the number of functions, as each function requires both a \textit{PACBTI} and a \textit{AUT} instruction. The only exception involves \textit{BTI} instructions, which are specifically added to manage indirect jumps.

\subsubsection{Comparison with similar technologies}

\gls{PACBTI} shares many similarities with \gls{CFI} solutions that employ a shadow stack to securely store a copy of the return address. Shadow stacks store a secure copy of the return addresses in a separate place in memory. These addresses are then used when a function returns to ensure the integrity of the return address. In \gls{PACBTI}, instead of storing the return address in a secure location, a signature of the return address is generated, which can be stored in an insecure location and later used to verify the return address before it is reused. 

Over the years, ARM Cortex-M processors have been largely overlooked in the development of advanced \gls{CFI} mechanisms. For example, while the LLVM compiler includes multiple \gls{CFI} implementations, such as shadow stacks, these features are not supported in this type of processor. Regardless, researchers have explored shadow stack implementations specifically tailored for Cortex M processors. Zhou et al.~\cite{Zhou2020} presented \textit{Silhouette}, a shadow stack implementation aimed at embedded systems,  and more recently Choi et al.~\cite{Choi2024} created \textit{SuM}, another  shadow stack implementation focused on ARM Cortex-M features. Both are compiler-based defenses implemented on top of the LLVM 9 compiler and require the existence of a \gls{MPU} to protect the shadow stack. 

The \textit{Silhouette} solution employs a parallel shadow stack~\cite{burow2019sok}, where the return addresses are stored at a fixed offset relative to their original location on the regular stack. The shadow stack is protected by the \gls{MPU}, which restricts access to privileged store instructions. To strengthen security, Silhouette ensures that software predominantly relies on unprivileged store instructions. Privileged store instructions are reserved for scenarios where their use is unavoidable, such as in the \gls{HAL} or when storing return addresses in protected memory regions.

The \textit{SuM}~\cite{Choi2024} distinguishes itself by using a \textit{FaultMask}, a flag that temporarily elevates the current execution priority, to enable access to the secure region where the shadow stack is stored, rather than replacing store instructions. Moreover, \textit{SuM} employs a compact shadow stack, storing the return address sequentially to minimize unused memory regions. The shadow stack's location is stored in a dedicated register to improve performance and is protected using the ARM Data Watchpoint Trace unit, which monitors any attempts to alter this register. 

Finally, both solutions analyze the code to identify indirect function calls and jumps, employing various techniques to minimize the risk of exploitation by attackers. \textit{SuM} introduces an additional safeguard by verifying whether the return address remains in a register throughout its lifecycle. Only if this condition is not met is the return address saved to the shadow stack, thereby eliminating unnecessary operations and enhancing efficiency.

Both solutions were tested by their authors using the same benchmarks we employed and, at first glance, achieved better results than ours in the BEEBS benchmark in terms of time overhead (Table \ref{tab:other-solutions-benchmark}). However,  since these values were not produced under the same conditions, they cannot be used for a direct comparison. Both solutions were implemented on top of the LLVM version 9, whereas \gls{PACBTI} was only introduced in LLVM version 17. Unfortunately, we were unable to integrate the two implementations into our environment to create a controlled setup and execute all benchmarks under identical conditions.

\begin{table}[]
\centering
 \resizebox{0.45\textwidth}{!}{\begin{tabular}{l|cc|cc}
\hline
\multicolumn{1}{c|}{\multirow{2}{*}{\textbf{Solutions}}} & \multicolumn{2}{c|}{\textbf{Execution overhead}} & \multicolumn{2}{c}{\textbf{Size Overhead}} \\ \cline{2-5} 
\multicolumn{1}{c|}{} & BEEBS  & CoreMark Pro & BEEBS   & CoreMark Pro \\ \hline
\textbf{Silhouette}   & 3.4\%  & 1.3\%        & 16.15\% & 14.42\%      \\
\textbf{SuM}          & 2.77\% & 2.63\%       & 8.83\%  & 6.59\%       \\
\textbf{PACBTI}       & 4.7\%  & 1.0\%        & 6.8\%   & 4.7\%       \\ \hline
\end{tabular}}
\caption{\label{tab:other-solutions-benchmark} Comparison between benchmark results with other solutions.}
\end{table}

Several observations can be made from the analysis. First, both solutions did not include the recursion test from the BEEBS benchmark in their evaluation, which was the test where our solution experienced the most significant overhead. Since \textit{Silhouette} and \textit{SuM} share similar operating principles with our solution, they likely introduced a similar increase in overhead in the BEEBS benchmark. If we calculate the geometric mean for this benchmark, excluding the recursion test, \gls{PACBTI} results in an overhead of 3.7\%. Furthermore, both solutions based their performance calculations on tests where their implementations executed faster than without them, while we disregarded such scenarios in our evaluation.

Regardless of the benchmark results, several important differences can influence the applicability of these solutions. First, solutions based on shadow stacks need additional memory regions, which can be a limitation in memory-constrained environments. In contrast, \gls{PACBTI} utilizes the \textit{r12} register to store the return address signature, which is eventually stored within the existing stack. As a result, while \gls{PACBTI} increases memory usage, it does not require separate memory regions and instead leverages the existing stack, making it more efficient in terms of memory usage. On the other hand, \textit{Silhouette} employs a parallel shadow stack that is the same size as the main stack, using the current stack pointer as an offset to store the return address. This approach imposes a strict requirement that may be challenging to meet on many resource-constrained devices.

In terms of increase in firmware size, \gls{PACBTI} is more lightweight than the other solutions (Table \ref{tab:other-solutions-benchmark}). This difference in size is not due to the core functionalities of the solutions, as both \gls{PACBTI} and a shadow stack implementation require the addition of four instructions to each function. Instead, it stems from the additional mechanisms implemented by the two shadow stack solutions. Both SuM and Silhouette~\cite{Zhou2020} incorporate mechanisms to control indirect jumps, with SuM introducing an extra mechanism to protect context switches between normal execution and exceptions by leveraging the shadow stack principle. Meanwhile, Silhouette converts privileged store instructions to unprivileged stores, which, in some cases, results in the addition of extra instructions~\cite{Zhou2020}.

In summary, while the precise overhead induced by \gls{PACBTI} compared to other solutions remains challenging to evaluate, a \gls{PACBTI}-based solution offers several advantages over shadow stack-based approaches, while providing a comparable level of protection against code reuse attacks. A solution based on \gls{PACBTI} does not require additional memory regions and is already supported by off-the-shelf compilers with minimal configuration. However, there is still potential for improvement in how compilers utilize this ARM extension. Currently, compilers add both \gls{PAC} and \gls{BTI} instructions to every function call, even when only \gls{BTI} is enabled, resulting in unnecessary landing pads for all functions, regardless of whether they are expected jump targets. Optimizations similar to those in SuM, where a control flow graph (\gls{CFG}) is computed during compilation to apply the shadow stack only in functions where it is genuinely required, could enhance both the performance and security of solutions based on this novel technology, thereby improving the effectiveness of solutions like RunPBA as well.

\section{Conclusion and Future Work}\label{conclusion}
In this work, we introduced RunPBA, a hardware-based runtime attestation system. It leverages the Arm PACBTI extension, an innovative technology found in certain processors within the Arm Cortex-M lineup. RunPBA achieves low overhead using off-the-shelf processors and is designed as an extension of the Trusted Firmware-M project, with the goal of facilitating easy integration into real-world systems. RunPBA is able to mitigate the \gls{TOCTOU} problem present in many attestation schemes by enforcing the intended execution flow. In addition to detecting runtime faults, RunPBA also offers the capability to trace them. We developed a prototype of the RunPBA solution, and our comprehensive evaluation shows that it introduces a minimal performance overhead, specifically 1\% on CoreMark Pro and 4.7\% on BEEBS. All code and tools used in this study are available at: \url{https://github.com/MrSuicideParrot/runpba}.

Furthermore, we envision potential avenues for future improvement. One promising direction is enhancing the compiler to better analyze source code and identify functions where PACBTI instructions may not contribute significantly to security, thereby optimizing runtime performance. Such improvements could benefit not only RunPBA but also any other solutions that leverage the PACBTI extension. We acknowledge the need to investigate new attack vectors, such as \gls{FOP} (Fault of Protection) attacks, which were outside the scope of this work but represent a critical area for future research in the context of runtime attestation systems. Finally, the potential of PACBTI to prevent data-flow attacks should be researched.

\section*{Declaration of Generative AI and AI-assisted technologies in the writing process}

During the preparation of this work, the authors used ChatGPT4 to revise the text throughout the paper, correcting typos and grammatical errors. After using this service, the authors reviewed and edited the content as needed and take full responsibility for the content of the publication.

\bibliographystyle{cas-model2-names}

\bibliography{cas-refs}

@InProceedings{Oh2017,
  author       = {Oh, Se-Ra and Kim, Young-Gab},
  booktitle    = {2017 International Conference on Platform Technology and Service (PlatCon)},
  title        = {Security requirements analysis for the IoT},
  year         = {2017},
  organization = {IEEE},
  pages        = {1--6},
  creationdate = {2025-06-06T10:52:39},
}

@Misc{TMP_threat_model,
  author       = {{TrustedFirmware-M Project}},
  howpublished = {\url{https://trustedfirmware-m.readthedocs.io/en/latest/security/threat_models/generic_threat_model.html}},
  note         = {Accessed: 2025-03-10},
  title        = {Generic Threat Model},
  creationdate = {2025-03-10T14:49:14},
}

@InProceedings{Shacham2007,
  author       = {Shacham, Hovav},
  booktitle    = {Proceedings of the 14th ACM Conference on Computer and Communications Security},
  title        = {The geometry of innocent flesh on the bone: return-into-libc without function calls (on the x86)},
  year         = {2007},
  address      = {New York, NY, USA},
  pages        = {552–561},
  publisher    = {Association for Computing Machinery},
  series       = {CCS '07},
  creationdate = {2024-04-02T15:42:00},
  doi          = {10.1145/1315245.1315313},
  isbn         = {9781595937032},
  location     = {Alexandria, Virginia, USA},
  numpages     = {10},
  url          = {https://doi.org/10.1145/1315245.1315313},
}

@InProceedings{Bletsch2011,
  author       = {Bletsch, Tyler and Jiang, Xuxian and Freeh, Vince W. and Liang, Zhenkai},
  booktitle    = {Proceedings of the 6th ACM Symposium on Information, Computer and Communications Security},
  title        = {Jump-oriented programming: a new class of code-reuse attack},
  year         = {2011},
  address      = {New York, NY, USA},
  pages        = {30–40},
  publisher    = {Association for Computing Machinery},
  series       = {ASIACCS '11},
  creationdate = {2024-04-02T15:43:10},
  doi          = {10.1145/1966913.1966919},
  isbn         = {9781450305648},
  location     = {Hong Kong, China},
  numpages     = {11},
  url          = {https://doi.org/10.1145/1966913.1966919},
}

@Misc{ArmLimited2024,
  author       = {{Arm Limited}},
  howpublished = {\url{https://tf-m-user-guide.trustedfirmware.org}},
  note         = {Accessed on June 7, 2024},
  title        = {{Trusted Firmware-M Documentation}},
  year         = {2024},
  creationdate = {2024-06-07T11:35:11},
}

@TechReport{ArmLimited2021,
  author       = {{Arm Limited}},
  title        = {Platform Security Model},
  year         = {2021},
  month        = dec,
  note         = {Accessed on June 13, 2024},
  number       = {JSADEN014},
  creationdate = {2024-06-13T17:19:57},
  howpublished = {\url{https://www.psacertified.org/app/uploads/2021/12/JSADEN014_PSA_Certified_SM_V1.1_BET0.pdf}},
}

@Article{Stratton2023,
  author       = {Stratton, Logan and Cronin, Kyle},
  title        = {Bypassing Modern CPU Protections With Function-Oriented Programming},
  year         = {2023},
  creationdate = {2024-03-08T16:46:09},
}

@Misc{ArmLimited2023a,
  author       = {{Arm Limited}},
  howpublished = {\url{https://developer.arm.com/documentation/ka005109/latest}},
  month        = nov,
  note         = {Accessed on May 22, 2023},
  title        = {PACMAN security vulnerability},
  year         = {2023},
  creationdate = {2024-05-21T11:56:09},
  url          = {https://developer.arm.com/documentation/ka005109/latest},
}

@InProceedings{Ravichandran2022,
  author    = {Ravichandran, Joseph and Na, Weon Taek and Lang, Jay and Yan, Mengjia},
  booktitle = {Proceedings of the 49th Annual International Symposium on Computer Architecture},
  title     = {PACMAN: Attacking ARM Pointer Authentication with Speculative Execution},
  year      = {2022},
  address   = {New York, NY, USA},
  pages     = {685–698},
  publisher = {Association for Computing Machinery},
  series    = {ISCA '22},
  doi       = {10.1145/3470496.3527429},
  isbn      = {9781450386104},
  location  = {New York, New York},
  numpages  = {14},
  url       = {https://doi.org/10.1145/3470496.3527429},
}

@Misc{Mujumdar2021,
  author       = {Alan Mujumdar},
  month        = apr,
  note         = {{Accessed: Apr 23, 2023}},
  title        = {Armv8.1-M Pointer Authentication and Branch Target Identification Extension},
  year         = {2021},
  creationdate = {2024-04-23T15:42:26},
  url          = {https://community.arm.com/arm-community-blogs/b/architectures-and-processors-blog/posts/armv8-1-m-pointer-authentication-and-branch-target-identification-extension},
}

@Misc{QualcommTechnologies2017,
  author       = {{Qualcomm Technologies, Inc.}},
  howpublished = {\url{https://www.qualcomm.com/content/dam/qcomm-martech/dm-assets/documents/pointer-auth-v7.pdf}},
  month        = mar,
  note         = {Accessed: 2024-05-16},
  title        = {Pointer Authentication on ARMv8.3 Design and Analysis of the New Software Security Instructions},
  year         = {2017},
  creationdate = {2024-05-16T15:14:31},
  url          = {https://www.qualcomm.com/content/dam/qcomm-martech/dm-assets/documents/pointer-auth-v7.pdf},
}

@InProceedings{Guo2018,
  author       = {Guo, Yingjie and Chen, Liwei and Shi, Gang},
  booktitle    = {2018 IEEE Conference on Communications and Network Security (CNS)},
  title        = {Function-oriented programming: A new class of code reuse attack in c applications},
  year         = {2018},
  organization = {IEEE},
  pages        = {1--9},
  creationdate = {2024-04-30T15:12:21},
}

@misc{Renesas_Trust_Firmware_2024,
  title        = {Arm Trust Firmware M on RA8M1 - Renesas Engineering Community Forum},
  author       = {Renesas},
  howpublished = {\url{https://community.renesas.com/mcu-mpu/ra/f/forum/34852/arm-trust-firmware-m-on-ra8m1}},
  year         = 2024,
  note         = {Accessed: 2024-08-19}
}

@misc{Renesas_Trusted_Firmware_Sample_2024,
  title        = {Request for the Latest Trusted Firmware-M Sample Project - Renesas Engineering Community Forum},
  author       = {Renesas},
  howpublished = {\url{https://community.renesas.com/mcu-mpu/ra/f/forum/35029/request-for-the-latest-trusted-firmware-m-sample-project}},
  year         = 2024,
  note         = {Accessed: 2024-08-19}
}

@Article{Choi2024,
  author       = {Wonwoo Choi and Minjae Seo and Seongman Lee and Brent Byunghoon Kang},
  journal      = {Computers \& Security},
  title        = {SuM: Efficient shadow stack protection on ARM Cortex-M},
  year         = {2024},
  issn         = {0167-4048},
  pages        = {103568},
  volume       = {136},
  creationdate = {2024-10-21T15:25:10},
  doi          = {https://doi.org/10.1016/j.cose.2023.103568},
  url          = {https://www.sciencedirect.com/science/article/pii/S0167404823004789},
}

@InProceedings{Zhou2020,
  author       = {Zhou, Jie and Du, Yufei and Shen, Zhuojia and Ma, Lele and Criswell, John and Walls, Robert J},
  booktitle    = {29th USENIX Security Symposium (USENIX Security 20)},
  title        = {Silhouette: Efficient protected shadow stacks for embedded systems},
  year         = {2020},
  pages        = {1219--1236},
  creationdate = {2024-04-08T10:13:19},
}

@InProceedings{burow2019sok,
  author       = {Burow, Nathan and Zhang, Xinping and Payer, Mathias},
  booktitle    = {2019 IEEE Symposium on Security and Privacy (SP)},
  title        = {SoK: Shining light on shadow stacks},
  year         = {2019},
  organization = {IEEE},
  pages        = {985--999},
  creationdate = {2024-10-21T15:48:03},
}

@Misc{EEMBC_CoreMark_PRO_2024,
  author       = {EEMBC},
  howpublished = {\url{https://www.eembc.org/coremark-pro/}},
  note         = {Accessed: 2024-10-12},
  title        = {CoreMark-PRO - EEMBC Embedded Microprocessor Benchmark Consortium},
  year         = {2024},
  creationdate = {2024-10-15T09:22:20},
}

@Article{Pallister2013,
  author       = {Pallister, James and Hollis, Simon and Bennett, Jeremy},
  journal      = {arXiv preprint arXiv:1308.5174},
  title        = {BEEBS: Open benchmarks for energy measurements on embedded platforms},
  year         = {2013},
  creationdate = {2024-09-12T08:50:54},
}

@Article{pallister2015identifying,
  author       = {Pallister, James and Hollis, Simon J and Bennett, Jeremy},
  journal      = {The Computer Journal},
  title        = {Identifying compiler options to minimize energy consumption for embedded platforms},
  year         = {2015},
  number       = {1},
  pages        = {95--109},
  volume       = {58},
  creationdate = {2024-09-12T10:08:51},
  publisher    = {Oxford University Press},
}

@Misc{ArmLimited2022,
  author       = {{{Arm Limited}}},
  howpublished = {\url{https://arm-software.github.io/psa-api/attestation/1.0/overview/report.html\#initial-attestation-report}},
  note         = {Accessed: Aug 19, 2024},
  title        = {Initial Attestation Report - PSA Certified Attestation API 1.0},
  year         = {2022},
  creationdate = {2024-08-19T10:30:11},
}

@TechReport{Lundblade2024,
  author       = {Laurence Lundblade and Giridhar Mandyam and Jeremy O'Donoghue and Carl Wallace},
  title        = {{The Entity Attestation Token (EAT)}},
  year         = {2024},
  month        = aug,
  note         = {Work in Progress},
  number       = {draft-ietf-rats-eat-30},
  type         = {Internet-Draft},
  creationdate = {2024-08-19T10:24:29},
  day          = {2},
  pagetotal    = {101},
  publisher    = {Internet Engineering Task Force},
  school       = {Internet Engineering Task Force},
  url          = {https://datatracker.ietf.org/doc/draft-ietf-rats-eat/30/},
}

@Manual{ArmLimited2020,
  title        = {Arm Cortex-M85 Processor Technical Reference Manual},
  author       = {{Arm Limited}},
  edition      = {r1p1},
  month        = dec,
  note         = {Accessed on May 22, 2024},
  year         = {2020},
  creationdate = {2024-05-22T15:34:51},
  howpublished = {https://developer.arm.com/documentation/101924/0101/?lang=en},
}

@Misc{Brand2019,
  author       = {Brandon Azad},
  howpublished = {\url{https://googleprojectzero.blogspot.com/2019/02/examining-pointer-authentication-on.html}},
  month        = Feb,
  note         = {Accessed: 2024-05-15},
  title        = {Examining Pointer Authentication on the iPhone XS},
  year         = {2019},
  creationdate = {2024-05-15T16:40:19},
  url          = {https://googleprojectzero.blogspot.com/2019/02/examining-pointer-authentication-on.html},
}

@InProceedings{Liljestrand2019,
  author    = {Liljestrand, Hans and Nyman, Thomas and Wang, Kui and Perez, Carlos Chinea and Ekberg, Jan-Erik and Asokan, N},
  booktitle = {28th USENIX Security Symposium (USENIX Security 19)},
  title     = {$\{$PAC$\}$ it up: Towards pointer integrity using $\{$ARM$\}$ pointer authentication},
  year      = {2019},
  pages     = {177--194},
}

@InProceedings{Walls2019,
  author       = {Walls, Robert J and Brown, Nicholas F and Le Baron, Thomas and Shue, Craig A and Okhravi, Hamed and Ward, Bryan C},
  booktitle    = {31st Euromicro Conference on Real-Time Systems (ECRTS 2019)},
  title        = {Control-flow integrity for real-time embedded systems},
  year         = {2019},
  organization = {Schloss-Dagstuhl-Leibniz Zentrum f{\"u}r Informatik},
  creationdate = {2024-04-11T11:44:26},
}

@InProceedings{Davi2015,
  author       = {Davi, Lucas and Hanreich, Matthias and Paul, Debayan and Sadeghi, Ahmad-Reza and Koeberl, Patrick and Sullivan, Dean and Arias, Orlando and Jin, Yier},
  booktitle    = {Proceedings of the 52nd Annual Design Automation Conference},
  title        = {HAFIX: Hardware-assisted flow integrity extension},
  year         = {2015},
  pages        = {1--6},
  creationdate = {2024-04-11T11:26:36},
}

@Article{Geden2023,
  author       = {Geden, Munir and Rasmussen, Kasper},
  journal      = {IET Information Security},
  title        = {Hardware-assisted remote attestation design for critical embedded systems},
  year         = {2023},
  number       = {3},
  pages        = {518-533},
  volume       = {17},
  creationdate = {2024-03-21T10:07:22},
  doi          = {https://doi.org/10.1049/ise2.12113},
  eprint       = {https://ietresearch.onlinelibrary.wiley.com/doi/pdf/10.1049/ise2.12113},
  url          = {https://ietresearch.onlinelibrary.wiley.com/doi/abs/10.1049/ise2.12113},
}

@InProceedings{Arias2020,
  author       = {Arias, Orlando and Sullivan, Dean and Shan, Haoqi and Jin, Yier},
  booktitle    = {2020 IEEE International Symposium on Hardware Oriented Security and Trust (HOST)},
  title        = {LAHEL: Lightweight Attestation Hardening Embedded Devices using Macrocells},
  year         = {2020},
  month        = {Dec},
  pages        = {305-315},
  creationdate = {2024-04-12T10:58:57},
  doi          = {10.1109/HOST45689.2020.9300257},
}

@InProceedings{Dessouky2018,
  author       = {Ghada Dessouky and Tigist Abera and Ahmad Ibrahim and Ahmad-Reza Sadeghi},
  booktitle    = {Proceedings of the International Conference on Computer-Aided Design},
  title        = {LiteHAX: Lightweight Hardware-Assisted Attestation of Program Execution},
  year         = {2018},
  month        = {nov},
  pages        = {1-8},
  publisher    = {{ACM}},
  creationdate = {2024-03-14T11:15:45},
  doi          = {10.1145/3240765.3240821},
  readstatus   = {read},
}

@InProceedings{Dessouky2017,
  author    = {Dessouky, Ghada and Zeitouni, Shaza and Nyman, Thomas and Paverd, Andrew and Davi, Lucas and Koeberl, Patrick and Asokan, N. and Sadeghi, Ahmad-Reza},
  booktitle = {Proceedings of the 54th Annual Design Automation Conference 2017},
  title     = {LO-FAT: Low-Overhead Control Flow ATtestation in Hardware},
  year      = {2017},
  address   = {New York, NY, USA},
  publisher = {Association for Computing Machinery},
  series    = {DAC '17},
  articleno = {24},
  doi       = {10.1145/3061639.3062276},
  isbn      = {9781450349277},
  location  = {Austin, TX, USA},
  numpages  = {6},
  url       = {https://doi.org/10.1145/3061639.3062276},
}

@InProceedings{nasahl2021protecting,
  author       = {Nasahl, Pascal and Schilling, Robert and Mangard, Stefan},
  booktitle    = {2021 IEEE International Symposium on Hardware Oriented Security and Trust (HOST)},
  title        = {Protecting Indirect Branches Against Fault Attacks Using ARM Pointer Authentication},
  year         = {2021},
  organization = {IEEE},
  pages        = {68--79},
  creationdate = {2024-10-30T08:51:35},
}

@Article{fanti2022toward,
  author       = {Fanti, Andrea and Perez, Carlos Chinea and Denis-Courmont, Remi and Roascio, Gianluca and Ekberg, Jan-Erik},
  journal      = {IEEE Transactions on Computer-Aided Design of Integrated Circuits and Systems},
  title        = {Toward register spilling security using LLVM and ARM pointer authentication},
  year         = {2022},
  number       = {11},
  pages        = {3757--3766},
  volume       = {41},
  creationdate = {2024-10-30T08:52:50},
  publisher    = {IEEE},
}

@Misc{liljestrand2020pacstackauthenticatedstack,
  author        = {Hans Liljestrand and Thomas Nyman and Lachlan J. Gunn and Jan-Erik Ekberg and N. Asokan},
  title         = {PACStack: an Authenticated Call Stack},
  year          = {2020},
  archiveprefix = {arXiv},
  creationdate  = {2024-09-23T10:56:05},
  eprint        = {1905.10242},
  primaryclass  = {cs.CR},
  url           = {https://arxiv.org/abs/1905.10242},
}

@InProceedings{denis2020camouflage,
  author       = {Denis-Courmont, R{\'e}mi and Liljestrand, Hans and Chinea, Carlos and Ekberg, Jan-Erik},
  booktitle    = {2020 57th ACM/IEEE Design Automation Conference (DAC)},
  title        = {Camouflage: Hardware-assisted cfi for the arm linux kernel},
  year         = {2020},
  organization = {IEEE},
  pages        = {1--6},
  creationdate = {2024-10-29T14:42:40},
}

@InProceedings{Davi2009,
  author       = {Davi, Lucas and Sadeghi, Ahmad-Reza and Winandy, Marcel},
  booktitle    = {Proceedings of the 2009 ACM workshop on Scalable trusted computing},
  title        = {Dynamic integrity measurement and attestation: towards defense against return-oriented programming attacks},
  year         = {2009},
  pages        = {49--54},
  creationdate = {2024-03-14T10:52:48},
}

@Article{Kuang2020,
  author       = {Boyu Kuang and Anmin Fu and Lu Zhou and Willy Susilo and Yuqing Zhang},
  journal      = {Computers \& Security},
  title        = {DO-RA: Data-oriented runtime attestation for IoT devices},
  year         = {2020},
  issn         = {0167-4048},
  pages        = {101945},
  volume       = {97},
  creationdate = {2024-03-19T17:42:38},
  doi          = {https://doi.org/10.1016/j.cose.2020.101945},
  url          = {https://www.sciencedirect.com/science/article/pii/S0167404820302212},
}

@InProceedings{hristozov2018practical,
  author    = {Hristozov, Stefan and Heyszl, Johann and Wagner, Steffen and Sigl, Georg},
  booktitle = {NDSS Workshop on Decentralized IoT Security and Standards (DISS)},
  title     = {Practical runtime attestation for tiny IoT devices},
  year      = {2018},
  volume    = {18},
}

@InProceedings{Huo2020,
  author       = {Huo, Dongdong and Wang, Yu and Liu, Chao and Li, Mingxuan and Wang, Yazhe and Xu, Zhen},
  booktitle    = {2020 IEEE 22nd International Conference on High Performance Computing and Communications; IEEE 18th International Conference on Smart City; IEEE 6th International Conference on Data Science and Systems (HPCC/SmartCity/DSS)},
  title        = {LAPE: A Lightweight Attestation of Program Execution Scheme for Bare-Metal Systems},
  year         = {2020},
  pages        = {78-86},
  creationdate = {2024-04-12T10:35:13},
  doi          = {10.1109/HPCC-SmartCity-DSS50907.2020.00011},
}

@InProceedings{Wang2023,
  author       = {Wang, Jinwen and Wang, Yujie and Li, Ao and Xiao, Yang and Zhang, Ruide and Lou, Wenjing and Hou, Y Thomas and Zhang, Ning},
  booktitle    = {32nd USENIX Security Symposium (USENIX Security 23)},
  title        = {$\{$ARI$\}$: Attestation of Real-time Mission Execution Integrity},
  year         = {2023},
  pages        = {2761--2778},
  creationdate = {2024-03-13T11:35:48},
}

@TechReport{Almakhdhub2020,
  author       = {Almakhdhub, Naif and Clements, Abraham Anthony and Bagchi, Saurabh and Payer, Mathias},
  title        = {uRAI: Return Address Integrity for Embedded Systems.},
  year         = {2020},
  creationdate = {2024-04-12T09:21:21},
  doi          = {10.14722/ndss.2020.24016},
  journal      = {Proceedings 2020 Network and Distributed System Security Symposium},
  publisher    = {Internet Society},
  school       = {Sandia National Lab.(SNL-NM), Albuquerque, NM (United States)},
}

@InProceedings{Nyman2017,
  author    = {Nyman, Thomas and Ekberg, Jan-Erik and Davi, Lucas and Asokan, N.},
  booktitle = {Research in Attacks, Intrusions, and Defenses},
  title     = {CFI CaRE: Hardware-Supported Call and Return Enforcement for Commercial Microcontrollers},
  year      = {2017},
  address   = {Cham},
  editor    = {Dacier, Marc and Bailey, Michael and Polychronakis, Michalis and Antonakakis, Manos},
  pages     = {259--284},
  publisher = {Springer International Publishing},
  isbn      = {978-3-319-66332-6},
}

@Article{Kawada2020,
  author       = {Kawada, Tomoaki and Honda, Shinya and Matsubara, Yutaka and Takada, Hiroaki},
  journal      = {International Journal of Parallel Programming},
  title        = {TZmCFI: RTOS-Aware Control-Flow Integrity Using TrustZone for Armv8-M},
  year         = {2020},
  issn         = {1573-7640},
  month        = jul,
  number       = {2},
  pages        = {216--236},
  volume       = {49},
  creationdate = {2024-04-12T09:45:41},
  doi          = {10.1007/s10766-020-00673-z},
  publisher    = {Springer Science and Business Media LLC},
}

@Article{Yeo2022,
  author       = {Yeo, Gisu and Kim, Yeryeong and Song, Suhyeon and Kwon, Donghyun},
  journal      = {IEEE Access},
  title        = {Efficient CFI Enforcement for Embedded Systems Using ARM TrustZone-M},
  year         = {2022},
  pages        = {132675-132684},
  volume       = {10},
  creationdate = {2024-04-11T16:51:17},
  doi          = {10.1109/ACCESS.2022.3230791},
}

@Misc{Nordic_Power_Profiler_2024,
  author       = {Nordic Semiconductor},
  howpublished = {\url{https://www.nordicsemi.com/Products/Development-hardware/Power-Profiler-Kit-2}},
  note         = {Accessed: 2024-09-12},
  title        = {Power Profiler Kit II - Nordic Semiconductor},
  year         = {2024},
  creationdate = {2024-09-12T11:30:25},
}

@InProceedings{shao2022faslr,
  author       = {Shao, Xinhui and Luo, Lan and Ling, Zhen and Yan, Huaiyu and Wei, Yumeng and Fu, Xinwen},
  booktitle    = {European Symposium on Research in Computer Security},
  title        = {fASLR: Function-based ASLR for resource-constrained IoT systems},
  year         = {2022},
  organization = {Springer},
  pages        = {531--548},
  creationdate = {2024-05-31T09:34:31},
}

@InProceedings{Zhang2021a,
  author       = {Zhang, Liqiang and Chen, Qingsong and Yan, Fei},
  booktitle    = {Applied Cryptography in Computer and Communications: First EAI International Conference, AC3 2021, Virtual Event, May 15-16, 2021, Proceedings 1},
  title        = {A Security Enhanced Key Management Service for ARM Pointer Authentication},
  year         = {2021},
  organization = {Springer},
  pages        = {41--55},
  creationdate = {2024-05-17T09:15:34},
}

@Article{Kuang2022,
  author       = {Boyu Kuang and Anmin Fu and Willy Susilo and Shui Yu and Yansong Gao},
  journal      = {Computers \& Security},
  title        = {A survey of remote attestation in Internet of Things: Attacks, countermeasures, and prospects},
  year         = {2022},
  issn         = {0167-4048},
  pages        = {102498},
  volume       = {112},
  creationdate = {2024-03-21T10:14:09},
  doi          = {https://doi.org/10.1016/j.cose.2021.102498},
  url          = {https://www.sciencedirect.com/science/article/pii/S0167404821003229},
}

@Article{Ankergaard2021,
  author    = {Ankerg{\aa}rd, Sigurd Frej Joel J{\o}rgensen and Dushku, Edlira and Dragoni, Nicola},
  journal   = {Sensors},
  title     = {State-of-the-art software-based remote attestation: Opportunities and open issues for Internet of Things},
  year      = {2021},
  number    = {5},
  pages     = {1598},
  volume    = {21},
  publisher = {MDPI},
}

@InProceedings{Abera2016,
  author       = {Tigist Abera and N. Asokan and Lucas Davi and Farinaz Koushanfar and Andrew Paverd and Ahmad-Reza Sadeghi and Gene Tsudik},
  booktitle    = {Proceedings of the 53rd Annual Design Automation Conference},
  title        = {Invited - Things, trouble, trust},
  year         = {2016},
  address      = {New York, NY, USA},
  month        = {jun},
  publisher    = {{ACM}},
  series       = {DAC '16},
  articleno    = {121},
  creationdate = {2024-02-28T16:35:39},
  doi          = {10.1145/2897937.2905020},
  isbn         = {9781450342360},
  location     = {Austin, Texas},
  numpages     = {6},
  readstatus   = {read},
  url          = {https://doi.org/10.1145/2897937.2905020},
}

@Article{Burow2017,
  author       = {Burow, Nathan and Carr, Scott A. and Nash, Joseph and Larsen, Per and Franz, Michael and Brunthaler, Stefan and Payer, Mathias},
  journal      = {ACM Comput. Surv.},
  title        = {Control-Flow Integrity: Precision, Security, and Performance},
  year         = {2017},
  issn         = {0360-0300},
  month        = {apr},
  number       = {1},
  volume       = {50},
  address      = {New York, NY, USA},
  articleno    = {16},
  creationdate = {2024-03-13T16:53:14},
  doi          = {10.1145/3054924},
  issue_date   = {January 2018},
  numpages     = {33},
  publisher    = {Association for Computing Machinery},
  url          = {https://doi.org/10.1145/3054924},
}

@InProceedings{Dang2015,
  author       = {Dang, Thurston HY and Maniatis, Petros and Wagner, David},
  booktitle    = {Proceedings of the 10th ACM Symposium on Information, Computer and Communications Security},
  title        = {The performance cost of shadow stacks and stack canaries},
  year         = {2015},
  pages        = {555--566},
  creationdate = {2024-04-08T08:48:52},
}

@InProceedings{ammar2024bridging,
  author       = {Ammar, Mahmoud and Abdelraoof, Ahmed and Vlasceanu, Silviu},
  booktitle    = {33rd USENIX Security Symposium (USENIX Security 24)},
  title        = {On Bridging the Gap between Control Flow Integrity and Attestation Schemes},
  year         = {2024},
  pages        = {6633--6650},
  creationdate = {2024-11-08T14:40:40},
}

@Article{Makhdoom2023,
  author       = {Imran Makhdoom and Mehran Abolhasan and Daniel Franklin and Justin Lipman and Christian Zimmermann and Massimo Piccardi and Negin Shariati},
  journal      = {{Computers \& Security}},
  title        = {Detecting compromised IoT devices: Existing techniques, challenges, and a way forward},
  year         = {2023},
  issn         = {0167-4048},
  pages        = {103384},
  volume       = {132},
  creationdate = {2024-03-13T16:17:45},
  doi          = {https://doi.org/10.1016/j.cose.2023.103384},
  url          = {https://www.sciencedirect.com/science/article/pii/S0167404823002948},
}

@Misc{LMS572023,
  author       = {LMS57},
  howpublished = {\url{https://github.com/LMS57/FOP_Mythoclast}},
  note         = {Accessed: January 6, 2025},
  title        = {FOP Mythoclast},
  year         = {2023},
  creationdate = {2025-01-06T11:40:48},
}

@Misc{Tschofenig2024,
  author       = {Hannes Tschofenig and Simon Frost and Mathias Brossard and Adrian L. Shaw and Thomas Fossati},
  howpublished = {IETF Datatracker},
  month        = {September},
  note         = {Accessed: March 17, 2025},
  title        = {{Arm's Platform Security Architecture (PSA) Attestation Token}},
  year         = {2024},
  creationdate = {2025-03-17T16:50:59},
  day          = {23},
  number       = {draft-tschofenig-rats-psa-token-24},
  school       = {IETF},
  series       = {Internet-Draft},
  type         = {Draft},
  url          = {https://datatracker.ietf.org/doc/draft-tschofenig-rats-psa-token/},
}

\appendix
\section{BTI-enabled compilation}\label{append:btifeature}

During the evaluation phase of RunPBA, we analyzed firmware compiled with different \gls{PACBTI} settings and compilers. Our goal was to understand in which situations the compiler would add a \gls{BTI} landing pad, so we used a decompiler to analyse the produced assembly and understand if the \gls{BTI} landing pad was only being added to functions that were being called by indirect reference. From this analysis, every compiler we observed that we tested added BTI landing pads to every function, including the ones that were only called by direct jumps. This behavior was observed when compiling the firmware with only \gls{BTI} enabled. The following list enumerates the compilers that we tested and observed this condition. 
\begin{itemize}
    \item LLVM Embedded Toolchain for Arm 17.0.0.1
    \item Arm C Compiler for Embedded 6.20
    \item Arm GNU Toolchain 13.2.Rel1-x86\_64-arm-none-eabi
\end{itemize}

The binaries we analyzed are available in our github repository: \url{https://github.com/MrSuicideParrot/pacbti-benchmark}.

\printcredits

\iffalse
\bio{}
Author biography without author photo.
Author biography. Author biography. Author biography.
Author biography. Author biography. Author biography.
Author biography. Author biography. Author biography.
Author biography. Author biography. Author biography.
Author biography. Author biography. Author biography.
Author biography. Author biography. Author biography.
Author biography. Author biography. Author biography.
Author biography. Author biography. Author biography.
Author biography. Author biography. Author biography.
\endbio

\bio{}
Author biography with author photo.
Author biography. Author biography. Author biography.
Author biography. Author biography. Author biography.
Author biography. Author biography. Author biography.
Author biography. Author biography. Author biography.
Author biography. Author biography. Author biography.
Author biography. Author biography. Author biography.
Author biography. Author biography. Author biography.
Author biography. Author biography. Author biography.
Author biography. Author biography. Author biography.
\endbio

\bio{}
Author biography with author photo.
Author biography. Author biography. Author biography.
Author biography. Author biography. Author biography.
Author biography. Author biography. Author biography.
Author biography. Author biography. Author biography.
\endbio
\fi

\end{document}